\pdfoutput=1

\let\mypdfximage\pdfximage
\def\pdfximage{\immediate\mypdfximage}

\documentclass[letterpaper, 11pt]{article}

\usepackage{amsmath}
\usepackage{amssymb}
\usepackage{amsthm}
\usepackage[font=small]{caption} \usepackage{epsfig}
\usepackage{geometry}
\usepackage{graphicx}
\usepackage{url}
\usepackage[colorlinks=true, citecolor=blue, filecolor=black, linkcolor=red, urlcolor=blue]{hyperref}
\usepackage{multirow}
\usepackage{makecell}
\usepackage{algorithm}
\usepackage{algpseudocode}
\usepackage[caption=false]{subfig}
\usepackage{soul} \setstcolor{red}
\setulcolor{blue}
\algdef{SE}[SUBALG]{Indent}{EndIndent}{}{\algorithmicend\ }%
\algtext*{Indent}
\algtext*{EndIndent}
\usepackage{rotating}
\usepackage{longtable}

\usepackage[capitalise,nameinlink,noabbrev]{cleveref} 

\author{Jiayuan Dong\footnote{\href{mailto:jiayuand@umich.edu}{jiayuand@umich.edu}, Ph.D. student, Mechanical Engineering, University of Michigan, Ann Arbor, MI 48109.},
Jiankan Liao\footnote{\href{mailto:jkliao@umich.edu}{jkliao@umich.edu}, Ph.D. student, Mechanical Engineering, University of Michigan, Ann Arbor, MI 48109.},
Xun Huan\footnote{\href{mailto:xhuan@umich.edu}{xhuan@umich.edu}, Assistant Professor, Mechanical Engineering, University of Michigan, Ann Arbor, MI 48109. \href{https://uq.engin.umich.edu}{https://uq.engin.umich.edu}}, and
Daniel Cooper\footnote{Corresponding author: \href{mailto:drcooper@umich.edu}{drcooper@umich.edu}, Assistant Professor, Mechanical Engineering, University of Michigan, Ann Arbor, MI 48109.}
}

\date{}

\renewcommand{\(}{\left(}
\renewcommand{\)}{\right)}

\newcommand{\vvvert}{|\kern-1pt|\kern-1pt|}

\newcommand{\II}{\mathbb{I}}

\newcommand{\PP}{\mathbb{P}}
\newcommand{\RR}{\mathbb{R}}

\newcommand{\CD}{\mathcal{D}}

\newcommand{\argmin}{\operatornamewithlimits{argmin}}

\crefname{section}{Sec.}{Sec.}
\Crefname{section}{Section}{Sections}
\crefname{subsection}{Sec.}{Sec.}
\Crefname{subsection}{Section}{Sections}
\crefname{figure}{Fig.}{Fig.}
\Crefname{figure}{Figure}{Figures}
\crefname{equation}{Eqn.}{Eqn.}
\Crefname{equation}{Equation}{Equations}

\crefdefaultlabelformat{#2\textup{#1}#3}

\title{Expert Elicitation and Data Noise Learning for Material Flow Analysis using Bayesian Inference}

\begin{document}

\maketitle

\begin{abstract} 
Bayesian inference allows the transparent communication of uncertainty in material flow analyses (MFAs), and a systematic update of uncertainty as new data become available. However, the method is undermined by the difficultly of defining proper priors for the MFA parameters and quantifying the noise in the collected data. We start to address these issues by first deriving and implementing an expert elicitation 
procedure suitable for generating MFA parameter priors. Second, we propose to learn the data noise concurrent with the parametric uncertainty. These methods are demonstrated using a case study on the 2012 U.S. steel flow. Eight experts are interviewed to elicit distributions on steel flow uncertainty from raw materials to intermediate goods. The experts' distributions are combined and weighted according to the expertise demonstrated in response to seeding questions. These aggregated distributions form our model parameters' prior. A sensible, weakly-informative prior is also adopted for learning the data noise. Bayesian inference is then performed to update the parametric and data noise uncertainty given MFA data collected from the United States Geological Survey (USGS) and the World Steel Association (WSA).The results show a reduction in MFA parametric uncertainty when incorporating the collected data. Only a modest reduction in data noise uncertainty was observed; however, greater reductions were achieved when using data from multiple years in the inference. These methods generate transparent MFA and data noise uncertainties learned from data rather than pre-assumed data noise levels,
providing a more robust basis for decision-making that affects the system.

    
\end{abstract} 

\section{Introduction}

Material flow analysis (MFA) is a foundational tool of industrial ecology research  \cite{bookBrunner}
and characterizes how a given material is transported and transformed through a supply chain. MFAs are key to identifying potential resource efficiency improvements (e.g., increased recycling), and to evaluating the upstream and downstream system impacts of local interventions; e.g., the potential to reduce greenhouse gas (GHG) emissions released during material production by improving downstream manufacturing process yields \cite{Cullen22}. MFAs have been used to help set environmental policies and goals by national governments (e.g., justifying Japan's reduce, reuse, and recycling laws), local governments (e.g., remedial action taken against toxic releases into New York City harbor), and companies (e.g., Toyota’s corporate MFA was used to set company goals for emissions and recycling)~\cite{articleGraedel}. 
The proliferation of MFA, however, is hindered by at least two major challenges. First is the long timeline for creating and updating detailed MFAs, currently taking months or even years. 
Second is the lack of uncertainty quantification (UQ) in most MFA results—a lack of UQ limits insight into the impacts, risks and unintended consequences of system interventions. It is increasingly accepted that UQ must be included in MFA results if they are to meaningful support informed decision- and policy-making \cite{articleGraedel, Schwab}. Bayesian methods help address these challenges of UQ and laboriousness in MFA, as explained below.

\subsection{Previous work on Bayesian inference in MFA}

Bayesian inference is a probabilistic approach to achieve data reconciliation that adjusts MFA variable estimates by combining prior knowledge with collected material flow data \cite{jaynes_2003, bertsekas_introduction_2008, Berger:1327974, RevModPhys.83.943}. The prior information is typically a combination of fact-based knowledge with subjective impressions based on experience \cite{MoyeLA}. The prior belief about an MFA variable, such as the mass flow between two processes in a factory, is expressed as a probability density function (PDF); e.g., a Gaussian PDF could be used to represent a prior belief that a mass flow is expected to be 10 $t$ with a variance of 1 $t^2$. MFA data are subsequently collected and the ``noise'' in the collected data---e.g., due to the error in a mass sensor reading---is also expressed as PDFs. 
In Bayesian inference, the collected MFA data are combined with the priors to generate an updated posterior belief, represented as updated conditional PDFs. 
 
Bayesian inference presents several advantages over other forms of MFA data reconciliation such as least squares optimization \cite{Kopec16, Yongxian}. First, it allows a rigorous quantification of uncertainty via the probability and statistics formalism, and allows flexible probability distributions able to capture high-order non-Gaussian and correlation effects. For example, the prior knowledge of an MFA variable might be best represented as a uniform rather than normal distribution if nothing is known other than the upper and lower bounds on the variable. Second, it is particularly suited for handling sparse, noisy, and incomplete data, and can incorporate multiple data streams simultaneously. Third, the Bayesian framework provides a natural entryway to inject domain knowledge, such as by using historical data or opinions from subject matter experts to form the prior distribution of the MFA variables \cite{Wang2003}. Bayesian inference can also be “chained” together, to perform sequential learning that iteratively assimilates new data as they become available.
Even when little data is available, the Bayesian approach can provide a practitioner with an MFA with associated uncertainties. 

Bayesian inference was first used in MFA in 2010 by Gottschalk \textit{et al.}~\cite{GOTTSCHALK2010320} to study nano-TiO$_2$ mass flows. They formed uniform and triangular prior distributions centered on values of historical data and performed Bayesian inference using a Metropolis sampling algorithm with simulated instead of measured data. In 2018, Lupton and Allwood~\cite{lupton2018} introduced additional MFA prior forms (e.g., Dirichlet priors) and conducted a case study on deriving the global steel flow. Their case study highlights some of the challenges of applying Bayesian inference to MFA: 

\begin{itemize}
\item \textbf{Assigning proper and rigorously justified prior distributions.} Lupton and Allwood's steel flow analysis \cite{lupton2018} used previous results from Cullen \textit{et al.}~\cite{CullenSteel} as the basis of their priors. However, historical data may not be always available and even when it is (and still deemed relevant) it remains unclear how to form a probability distribution that properly reflects the uncertainty. Assuming a prior variance without justification can introduce bias to the posterior results. 
Alternatively, non-informative or weakly-informative priors may be used; e.g., assigning a wide uniform PDF for a mass flow between 0 Mt and 200 Mt. However, they will likely require more MFA data to be collected in order to decrease uncertainty to desirable levels.  
    
\item \textbf{Assigning noise to collected MFA data.}
MFA relevant data are typically published without accompanying uncertainty information; e.g., no error bars are given for the commodity mass flow data reported by the U.N. Comtrade Database \cite{UNcomtrade12} or from trade associations such as the World Steel Association \cite{WorldSteel12}.
     
How then to model the data noise? Assumptions can be made; e.g., Lupton and Allwood~\cite{lupton2018} assume the data noise in their collected data follows a Gaussian distribution with a standard deviation equal to 10\% of the observed value. Such assumptions can introduce bias into the final posterior MFA results and provide a false sense of confidence if the assumed noise over-estimates the data 
quality. 
Elsewhere, there are qualitative methods to categorize data into uncertainty levels based on features such as the perceived data source quality and specificity \cite{bonnin2013}, and semi-quantitative approaches such as using a pedigree matrix or confidence score \cite{Lloyd2007, Yongxian} which translates an uncertainty level into a probability distribution or numerical value. The strict quantitative identification of MFA data uncertainty has been lacking so far \cite{Do2014}; however, the Bayesian framework can also facilitate the \emph{learning} of data noise.
\end{itemize}

\subsection{Scope and structure of this article}
This paper explores how to: (a) Form informative prior distributions for MFA variables by eliciting information from industrial experts, and (b) Learn the data noise from collected data by incorporating data noise as random variables for inference. In \cref{s:formulation}, we first introduce the MFA problem mathematically using the conservation of mass principle (\cref{ss:MFA_model}) and then formulate Bayesian inference for learning MFA model parameters and the collected data noise (\cref{ss:Bayes}). The two crucial components needed for solving a Bayesian inference problem---the likelihood and prior---are then presented in \cref{ss:likelihood,ss:elicitation} respectively. In particular, \cref{ss:elicitation} reviews existing methods for expert elicitation and discusses their use for the MFA framework. 
In \cref{s:USGS}, we apply these methods to derive the U.S. steel flow for 2012. Finally, in \cref{s:discussion} we discuss the lessons learned from the case study.

\section{Formulation}
\label{s:formulation}

\subsection{A Mathematical Representation of MFA}
\label{ss:MFA_model}

\begin{figure}[h]
	\centering
		\includegraphics[width=100mm]{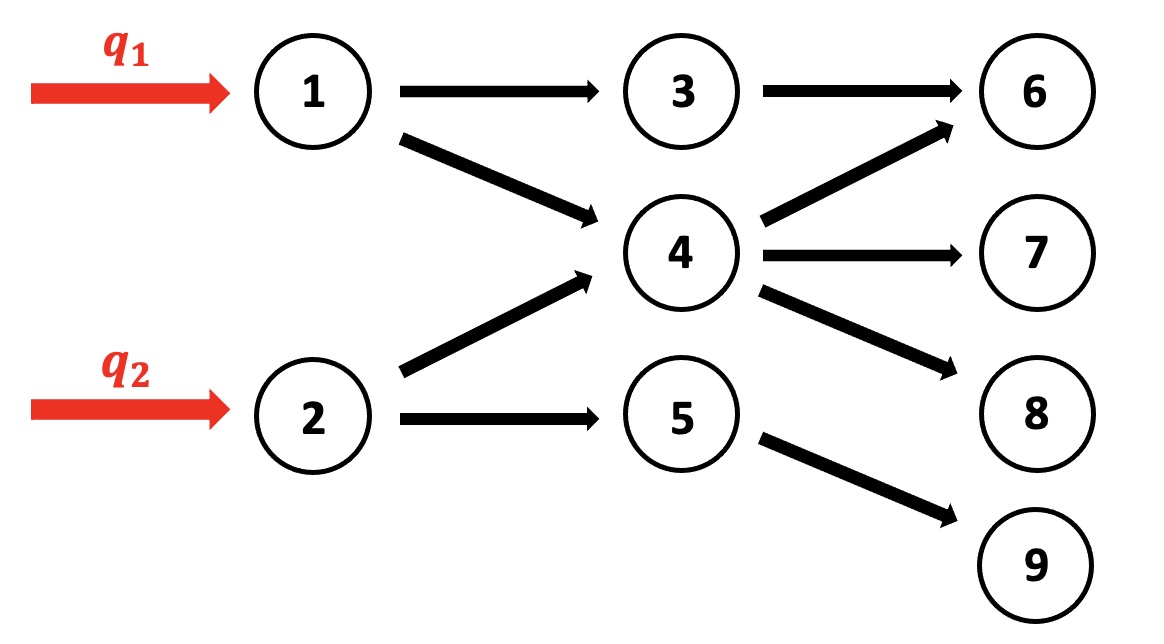}
  \caption{A graphical representation of an MFA network structure}
  \label{f:MFA_example}
\end{figure}

An MFA can be represented via a directed graph as shown in \cref{f:MFA_example}. The nodes of the graph (indexed by $1,2,\ldots, n_p$) represent different processes, products or locations. Each directed edge connecting two nodes represents the mass flow of material from one process to another.

At the core of MFA is the conservation of mass, which requires the total input flow of material mass for each node to equal its total output flow. We denote the total input (equivalently, total output) flow for node $i$ by $z_i$. The flow along an edge out of node $i$ (e.g., to node $j$) is then equal to $\phi_{ij} z_i$, where $\phi_{ij} \in [0,1]$ is the allocation fraction of node $i$'s total outflow going into node $j$ ($\phi_{ij}=0$ if there is no flow from node $i$ to node $j$). Hence, 
\begin{align}
     \sum_{i=1}^{n_p}\phi_{ij}z_{i}=z_{j}.\label{e:sum_flow}
\end{align}
Furthermore, for each node, its output allocation fractions need to sum to unity: 
\begin{align}
    \sum_{j=1}^{n_p}\phi_{ij} = 1. \label{e:sum_fractions}
\end{align}

We recommend working with the allocation fractions ($\phi_{ij}$) as model parameters instead of working directly with the mass flow values for each edge. The allocation fractions offer a convenient method of expressing and allowing the mass balance relationships for the entire MFA to be assembled into a linear system as proposed by Gottschalk \textit{et al.} \cite{Gottschalk2010}. For instance, the mass balance equations for the simple MFA shown in \cref{f:MFA_example} can be expressed as: 
\begin{align}
\underbrace{
\begin{bmatrix}
    1  &  0  &  0  &  0  &  0  &  0  &  0  &  0  &  0   \\
    0 & 1 & 0 & 0 & 0 & 0 & 0 & 0 & 0 \\
    -\phi_{13} & 0 & 1 & 0 & 0 & 0 & 0 & 0 & 0          \\
    -\phi_{14} & -\phi_{24} & 0 & 1 & 0 & 0 & 0 & 0 & 0 \\  
    0  &  -\phi_{25}  &  0  &  0  &  1  &  0  &  0  &  0  &  0   \\
    0 & 0 & -\phi_{36}  & -\phi_{46} & 0 & 1 & 0 & 0 & 0 \\
    0  &  0  &  0  &  -\phi_{47}  &  0  & 0  &  1  &  0  &  0\\
    0  &  0 &  0  &  -\phi_{48}  &  0  &  0  &  0  &  1  &  0\\
    0 & 0 & 0 & 0 &  -\phi_{59} & 0 & 0 & 0 & 1 \\
\end{bmatrix}}_{\II-\Phi^T}
\underbrace{
\begin{bmatrix}
    z_1  \\
    z_2 \\
    z_3  \\
    z_4 \\
    z_5  \\
    z_6 \\
    z_7 \\
    z_8  \\
    z_9\\
\end{bmatrix}}_{z}
=
\underbrace{
\begin{bmatrix}
    q_1  \\
    q_2 \\
    0  \\
    0  \\
    0 \\
    0  \\
    0  \\
    0 \\
    0  \\
\end{bmatrix}}_{q}
\label{e:linear_system}
\end{align}
where $\II$ is the $n_p\times n_p$ identity matrix, $\Phi \in \RR^{n_p\times n_p}$ is the adjacency matrix where entries are the allocation fractions $\phi_{ij}$, $z \in \RR^{n_p}$ depicts all the nodal mass flows, and $q\in \RR^{n_p}$ represents the external inflows to the network. The material inflows ($q_i$) originate from outside the network. For example, a material inflow to the U.S. aluminum material flow network could be imports of aluminum billet. 
 
For a scenario with given $\Phi$ and $q$, we can compute the model prediction for all nodal mass flows via: 
\begin{align}
    z = (\II-\Phi^T)^{-1}q.
    \label{e: matrix inverse}
\end{align}
Subsequently, from the values of $\{z, \Phi, q\}$, other common MFA quantities of interest (QoIs) can be predicted, such as mass flows for each edge  ($\phi_{ij}z_i$), and sums, products and ratios of mass flows. We express these QoIs as a function, $G(\phi, q)$, where $\phi$ denotes a flattened vector containing all $\phi_{ij}$'s.

\subsection{Bayesian Parameter Inference}
\label{ss:Bayes}

Given an MFA model,
the set of all unknown MFA parameters $\theta \in \RR^{n_{\theta}}$ and the collected MFA data $y \in \RR^{n_y}$ are treated as random variables and associated with a joint
PDF
$p(\theta,y)$.  
Here, we use $\theta$ to denote \emph{all} MFA model parameters we are interested to learn, which may encompass $\phi$ and $q$ as well as other parameters; $y$ is the flattened set of collected and noisy MFA data records that correspond to the model prediction QoIs $G(\phi,q)$.
Bayes' rule then directly follows from the axioms of probability, stating:
\begin{align}
    p(\theta|y) = \frac{p(\theta,y)}{p(y)} = \frac{p(y|\theta)p(\theta)}{p(y)},
    \label{e:Bayes}
\end{align}
where
$p(\theta)$ 
is the prior PDF representing the initial belief in the MFA parameters $\theta$ before having collected any MFA data; $p(y|\theta)$ is the likelihood PDF (see \cref{ss:likelihood}); $p(\theta|y)$ is the posterior PDF representing the updated belief in the MFA parameters $\theta$ after having collected the MFA data $y$; and $p(y)$ is the model evidence (marginal likelihood) and acts as a normalizing constant for the posterior PDF. Performing Bayesian parameter inference entails computing or characterizing the posterior $p(\theta|y)$ while accessing the likelihood and prior. 

\subsection{Modeling the Likelihood $p(y|\theta)$}
\label{ss:likelihood}

The likelihood computes the probability of having collected MFA data $y$ if the model parameters had the value equal to $\theta$; that is, it provides a probabilistic measure on the mismatch between observation $y$ and model prediction $G(\phi, q)$.
There are many ways in which the collected data $y$ may relate to $G(\phi, q)$. For example, the discrepancy may be viewed as an additive noise: 
\begin{align}
    y_k = G_k(\phi, q) + \epsilon_k, \label{e:additive_error}
\end{align}
where $k$ indicates the $k$th data component. 
\Cref{e:additive_error} is appropriate for MFA data where the error is insensitive to the scale of the measurement; e.g., in the case of a mass sensor which has a sensitivity of $\pm$ 10 grams across its measurement range. However, oftentimes the data error increases with the scale of the measurement and can be modeled as a relative noise in the form: 
\begin{align}
    y_k = G_k(\phi, q)(1 + \epsilon_k). \label{e:relative_error}
\end{align}
In either case, modeling the noise $\epsilon_k$ as a Gaussian centered around zero is a sensible assumption. For its standard deviation $\sigma_k$, we do not presume its magnitude and will instead learn these values from the data. In the absence of information to the contrary, the noise associated with different pieces of data can be modeled as independent. Therefore, if adopting the relative error form in \cref{e:relative_error} and with the unknown model parameters $\theta=\{q, \phi,\sigma\}$, the likelihood PDF is: 
\begin{align}
p(y|\theta) = p_{\epsilon} \left(\frac{y}{G(\phi, q)}-1\right) =  \prod_{k=1}^{n_y}\frac{1}{\sqrt{2\pi} \sigma_{k}}\exp\left\{-\frac{\(\frac{y_k}{G_k(\phi, q)}-1\)^2}{2\sigma_k^2}\right\}.
\label{e:the_likelihood}
\end{align}

\subsection{Expert Prior Elicitation for MFA}
\label{ss:elicitation}

The goal of expert prior elicitation is to extract pertinent knowledge for $\theta$ from subject matter experts in a form that can used as a proper Bayesian prior PDF. There are two main challenges in expert prior elicitation. An expert does not typically have a preexisting quantification of her belief in the form of a PDF~\cite{Winkler}. Therefore, the first challenge is how to elicit and synthesize a single expert's knowledge into a ``quantified belief prior''~\cite{Winkler}. Next, when there are multiple experts available, it is desirable to utilize all their opinions to have the prior capture the full diversity of background knowledge. However, Bayesian inference requires a single PDF for a parameter as the prior rather than multiple distributions from several experts. The second challenge is, therefore, how to combine and weight the beliefs of multiple experts into a single prior for each MFA variable of interest. 

Expert prior elicitation has been widely applied in medicine to design clinical trials \cite{Azzolina2021}, assess the effect of specific treatments for rare disorders \cite{Ramanan2019}, and inform health-care decision-making \cite{Bojke21}. Prior elicitation has also been used in environmental assessments to forecast future wind energy costs \cite{Wiser2016} and regional climate change \cite{Dessai_2018}. There has been some preliminary work on expert elicitation in MFA:  Montangero and Belevi \cite{MONTANGERO20071052} use expert elicitation to describe the uncertainty regarding the flow of nitrogen and phosphorus in a septic tank. In their work, multiple experts provide uncertainty information as quantiles and their opinions are combined using equal weights. Elsewhere, Mathieux and Brissaud \cite{MATHIEUX2010} conduct expert elicitation to understand where aluminum is being used in commercial vehicles. In their work, experts are brought together to determine collectively a single distribution for each estimate. In this work, we examine methods for eliciting, weighting, and aggregating multiple experts' beliefs under the Bayesian framework. 

\subsubsection{Eliciting a Prior from an Expert}
Prior elicitation is typically performed using surveys conducted either remotely (e.g., by mail or online), in-person, or via a hybrid format where the expert is assisted by telephone or video conference \cite{Johnson2010MethodsTE}. While the variable of interest $\theta$ can be multivariate (e.g., a joint distribution on all the allocation fractions leaving a node), eliciting multi-dimensional PDFs directly is very challenging; therefore, multivariate elicitation typically involves eliciting and then combining univariate marginal distributions \cite{daneshkhah2010eliciting, OHagan2006a}. The common methods to elicit univariate PDFs are either a \textbf{Variable Interval Method} or a \textbf{Fixed Interval Method} \cite{OHagan2006a, oakley2010eliciting}. In a Variable Interval Method, the expert provides estimates of the quantiles; e.g., estimate $a$ and $b$ such that $\PP(\theta \leq a) = 0.25$, $\PP(a < \theta \leq  b) = 0.5$ and  $\PP(b > \theta ) = 0.25$ \cite{MurphyandWinkler1974, GarthwaiteOhagan}. In Fixed Interval Methods, the expert estimates the probability of $\theta$ within given fixed intervals (e.g., estimate $\PP(a < \theta \leq b)$ for some given ${a}$ and $b$) \cite{OHagan1998}. While it is unclear which set of methods yield more accurate representations of the expert's true belief (Abbas' findings contradicting those of Murphy and Winkler \cite{Abbas08,MurphyandWinkler1974}) it does appear that participants find the Fixed Interval Method easier to complete \cite{Abbas08}. Given that many MFA industry experts might be unfamiliar with statistical concepts such as quantiles, we recommend using the Fixed Interval Method for MFA prior elicitation where possible; e.g., for eliciting allocation fractions ($\phi$) which are bounded within $[0, 1]$. Elsewhere, when eliciting external inflows ($q$) or data noise parameters ($\sigma_k$), there is no upper bound \textit{a priori} so that it is appropriate to either ask the expert to specify an upper bound 
before defining the intervals used in the Fixed Interval Method or else use a Variable Interval Method (see SI Section 1).  %

\subsubsection{Prior Aggregation from Multiple Experts}

The typical methods for combining multiple experts' knowledge into a single proper prior PDF are behavioral aggregation and mathematical aggregation \cite{OHagan2006a}. 

\paragraph{Behavioral Aggregation:} 
Experts collaborate to define agreed upon priors~\cite{OHagan2006a}. Thus, very `informed' experts have the chance to share their knowledge. However, it can be difficult to find a common time when all the experts are available and there are potential issues with strong personalities dominating the decision-making~\cite{OHagan2006a}, the risk of overconfidence in the group~\cite{Gigone1993}, some experts concealing their true views~\cite{Plous, SNIEZEK1992124}, group polarisation~\cite{Plous, SNIEZEK1992124}, and individuals with unique information being ineffective at sharing~\cite{Stasser1985PoolingOU}. Therefore, methods such as the Delphi method~\cite{bookLinstone,ROWE1999353} and its variants \cite{Degroot1974, Delbecq75} have been proposed where direct interaction between the experts is restricted to prompt the experts to explain their views rather than relying on reputation or personality~\cite{OHagan2006a}; e.g., experts may share their views anonymously, adjust their views based on the received information, and iterate until convergence on a distribution. 

\begin{figure}[h]
 	\centering
 	\includegraphics[width=85mm]{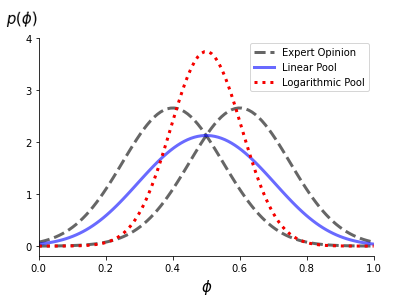}
     \caption{Linear versus logarithmic pooling of two equal weight elicited priors for an allocation fraction ($\phi$). In general, the logrithmic pool result is more ``concentrated'' than the linear pool result  \cite{OHagan2006a}.
     }
    \label{f: pool}
 \end{figure}

\paragraph{Mathematical Aggregation:}
A distribution is elicited from each of the $n_e$ experts independently, yielding $n_e$ PDFs $\{p_1(\theta), \ldots ,p_{n_e}(\theta) \}$. 
An aggregated PDF $p(\theta)$ is then calculated using either linear or logarithmic pooling~\cite{OHagan2006a, genest1986combining,Clemen99}: 
\begin{align}
\mathrm{(linear)\hspace{2em}} p(\theta) = \sum_{\ell=1}^{n_e}w_{\ell} p_{\ell}(\theta),
\hspace{6em} \mathrm{(logarithmic)\hspace{2em}} p(\theta) = \frac{1}{Z}\prod_{\ell=1}^{n_e}p_{\ell}(\theta)^{w_{\ell}},\label{e:pooling}
\end{align}
where $w_{\ell}$ is the weight associated with expert ${\ell}$, and $Z$ is a normalization constant. In logarithmic pooling, the resulting prior $p(\theta) = 0$ whenever any of the experts believe $p_{{\ell}}(\theta) = 0$. In contrast, the prior generated by linear pooling includes any value considered plausible by any of the experts and is therefore more conservative in terms of not ruling out experts' beliefs (see \cref{f: pool}).
While equal weights could be assigned to all experts such that $w_{\ell} = 1/n_e$, it is often desirable to allocate greater weighting to more informed experts, typically using either Social Network Weighting or methods requiring questions on \emph{seeding variables}, also referred to as \emph{seeding questions} from hereon. Social Network Weighting assigns weights based on each expert's number of citations \cite{COOKE2008} or a consensus among the experts on whose opinion should receive the most weight \cite{Aspinall2011}. Such methods have been criticized for often excluding experts with predominately industry rather than academic experience and for resulting in prior PDFs with low accuracy \cite{COOKE2008, Aspinall2011, Colson2018}. Seeding 
questions assess each expert's expertise by comparing the experts' responses to collected seeding variable observations. One seeding variable example for a steel MFA study might be on the fraction of U.S. pig iron consumed in Indiana within a certain time period. The experts' responses are compared to the observations from a credible source (e.g., USGS). Typically, Cooke's method \cite{Cooke91} is used to convert expert seeding question responses to expert weights where, for expert ${\ell}$, the weight $w_{\ell} \propto C_{\ell} K_{\ell}$ with $C_{\ell}$ being the calibration score and $K_{\ell}$ the information score. The calibration score measures the accuracy of the expert's responses, and the information score penalizes experts weakly informative responses that approach the uniform distribution. The Kullback-Leibler (KL) divergence is used in calculating both scores. The KL divergence is a measure of how two PDFs differ. It is non-symmetric and non-negative with a KL divergence being zero between two identical distributions, and a larger KL divergence implying a greater difference between two distributions \cite{kullback1997}. For discrete variable $X$ taking values in $\{1, 2, ..., m\}$, and two probability mass functions $P(x) = p_x$ and $Q(x) = q_x$, the KL divergence from $Q$ to $P$ is given as:
 \begin{align}
     \centering
    D_{\mathrm{KL}}(P||Q) = \sum_{i=1}^{m}p_i \ln \left(\frac{p_i}{q_i}\right). 
    \label{e:KLDivergence}
 \end{align}

\begin{figure}[htp]
 \centering
 \subfloat[Calculating the information score for an expert.]{\label{f:Info_Component}
 \includegraphics[width=140mm]{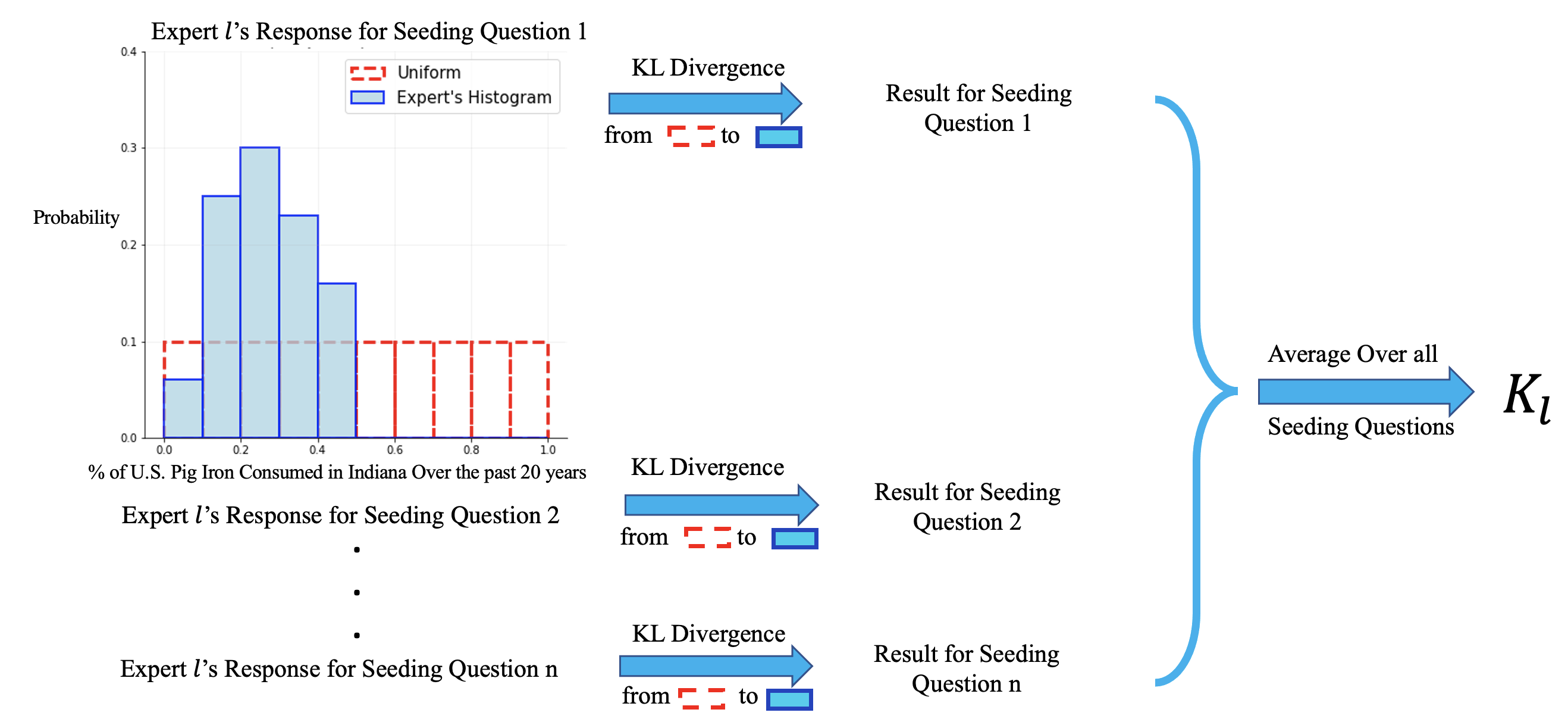}}\\
 \subfloat[Calculating the calibration score for an expert.]{\label{f:Cali_Component}\includegraphics[width=145mm]{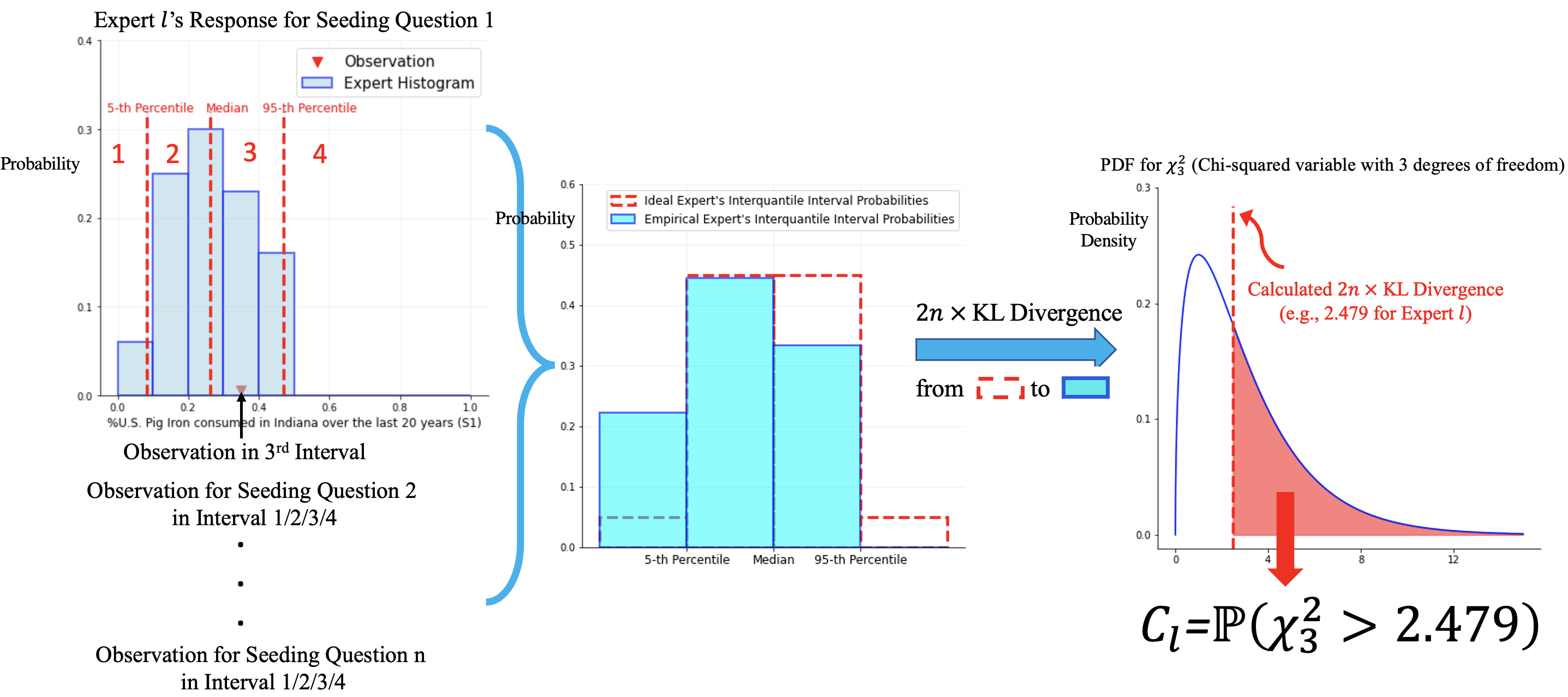}} 
 \caption{(a) Each expert's information score $K_{\ell}$ is calculated as the KL divergence from the expert's elicited histogram to the uniform distribution averaged across all seeding question responses. (b) Each expert's calibration score $C_{\ell}$ is calculated by first counting the number (fraction) of actual observations of the seeding questions
 in each inter-quantile interval across all seeding question responses from an expert,  then computing the KL divergence from  the ideal expert's inter-quantile interval probabilities to the empirical expert's inter-quantile interval probabilities, and lastly obtaining the likelihood ratio statistic with the corresponding $p$-value (i.e. $\mathbb{P}(\chi_3^2 > 2.479)$ in the above example) being the calibration score. }
    \label{Info_and_Cali_Component}
\end{figure}

Each expert's information score $K_{\ell}$ is calculated as the KL divergence from the expert's elicited distribution to the uniform distribution, averaged across all the seeding questions (see \cref{f:Info_Component}). In order to calculate an expert's calibration score $C_{\ell}$, each response to the seeding questions from expert ${\ell}$ is split into inter-quantile intervals. Typically, four inter-quantile intervals (3 degrees of freedom) are adopted with a corresponding probability vector $Q = \{0.05, 0.45, 0.45, 0.05\}$. Each seeding variable observation is then compared to the inter-quantile intervals derived from the expert's response; e.g., the USGS records that 34.9\% of U.S. pig iron was consumed in Indiana from 2002--2016, falling into the 50--95\% interval of the response from expert $\ell$ shown in \cref{f:Cali_Component}. %
Then, let $P_{\ell} = \{p_1, p_2, p_3, p_4\}$ denote the fraction of all $n$ seeding variable observations that fall into each of the four intervals elicited from expert ${\ell}$. Cooke \cite{Cooke91} states that for a well-informed ``ideal'' expert (i.e. where seeding variable observations
appear as independently drawn from a distribution consistent with the expert's quantiles) %
then $P_{\ell}$ tends to $Q$, and $D_{\mathrm{KL}}(P||Q)$ 
tends to zero. Cooke defines the calibration score $C_{\ell}$ as the probability that a random variable following a Chi-square distribution (with 3 degrees of freedom if using four inter-quantile intervals) is greater than the likelihood ratio statistic ($2 \times n \times D_{\mathrm{KL}}(P||Q)$), see \cref{f:Cali_Component}; therefore, if expert ${\ell}$'s knowledge differs from the seeding variable observations to a large extent, the associated KL divergence would be high and $C_{\ell}$ becomes small. While not discussed previously, we believe that Cooke's method of calculating $C_{\ell}$ is only appropriate when there is  uncertainty in the seeding variable observation
(see SI Section 2.4 for a detailed discussion).
Eggstaff \textit{et al.} state that there is no definitive minimum number of seeding questions; however, regardless of the number of questions, Cooke's method significantly outperforms equally weighting the experts' judgments \cite{Eggstaff2014}. 

{Once the expert weights are computed, they can then be used to carry out the mathematical aggregation in \cref{e:pooling} to complete the prior construction}. 

\subsection{Distributions for Modeling the MFA Priors}

The next step is fitting the histograms elicited from the MFA experts to a family of parameterized PDFs. The distribution and corresponding hyper-parameters are typically fitted and selected via a least squares procedure \cite{McBride2012, SHELF}.

\subsubsection{Distributions for Allocation Fraction Priors}
Lupton and Allwood \cite{lupton2018} proposed using a Dirichlet distribution
as the prior for the allocation fractions $\{ \phi_{S, d_1}, ..., \phi_{S,d_S} \}$
from source node $S$, ensuring that all the allocation fractions remain in %
$[0,1]$ and sum to unity. The Dirichlet PDF for $\phi_{S, d_1}, ..., \phi_{S,d_S}$ given hyper-parameters  $\alpha_D = \{\alpha_{d_1},...,\alpha_{d_S}\}$ is
\begin{align}
    p(\phi_{S, d_1}, ..., \phi_{S,d_S}|\alpha_D) = \frac{\Gamma\left(\sum_{i=d_1}^{d_S}\alpha_i\right)}{\prod_{j=d_1}^{d_S}\Gamma(\alpha_j)}\prod_{k=d_1}^{d_S}\phi_{S, k}^{\alpha_k-1}\label{e:Dirichlet}
\end{align}
and for each $\phi_{S,i}$, its marginal PDF follows a Beta distribution characterized by
\begin{align}
    p(\phi_{S,i}|\alpha_D) = \frac{\Gamma\left(\sum_{j=d_1}^{d_S}\alpha_j\right)}{\Gamma(\alpha_i)\Gamma\left(\sum_{k \neq i}\alpha_k\right)} \phi_{S,i}^{\alpha_i-1}(1-\phi_{S, i})^{(\sum_{l \neq i}\alpha_l) - 1}.\label{e:beta_marginal}
\end{align}

Another benefit of using a Dirichlet distribution is that eliciting each marginal distribution on $\phi_{S, i}$ (\cref{e:beta_marginal}) from an expert is sufficient to fully construct the  Dirichlet joint distribution (\cref{e:Dirichlet}) \cite{OHagan2006a}. After eliciting each marginal histogram of $\phi_{S, i}$ from the experts, we fit for the optimal Dirichlet hyper-parameters $\alpha_D^{\ast}$ by minimizing the squared differences between the probability of each Beta marginal that corresponds to the Dirichlet joint distribution and the marginal weighted histograms with $n_b$ intervals from the experts, summed across all $d_S$ allocation fractions emanating from a source node $S$ \cite{Zapata14}:
\begin{align}
    \alpha_{\CD}^{\ast} = \argmin_{\alpha_{\CD}} \sum_{d_s = d_1}^{d_S}\sum_{i=1}^{n_{b}} \left[F(\theta_{d_s, i+1}|\alpha_{\CD}) - F(\theta_{d_s, i}|\alpha_{\CD}) - \PP(\theta_{d_s, i})\right]^2,
    \label{e: least square fitting}
\end{align}
where $F$ is the Beta CDF.

Other methods exist that may also be used to model the allocation fraction priors \cite{lupton2018, Gelman1996}. For example, softmax transformations offer extra flexibility compared to using Dirichlet priors; e.g., they can incorporate a strong belief that $\phi_{S, d_1} = \phi_{S, d_2}$ while the relationship between other allocation fractions is unknown. However, the increased complexity of the procedure complicates the elicitation process because, for example, the CDF in \cref{e: least square fitting} cannot be evaluated in a closed-form.

\subsubsection{Distributions for Data Noise Parameter Priors $\sigma_k$}

Since $\sigma_k$ is the standard deviation of Gaussian distributed $\epsilon_k$, the prior must take positive support; i.e., $p(\sigma_k<0)=0$. A common choice of prior for the standard deviation of a Gaussian distribution is the Inverse-Gamma distribution, which enables an analytical evaluation of the posterior when the measured data is linear in the parameters of interest \cite{Gelman06}. However, the Inverse-Gamma and other commonly used positive support distributions such as the Log-Normal distribution place negligible probability of $\sigma$ at regions close to zero, greatly reducing
the possibility that the MFA data is of high quality.
We believe that generally in MFA there should be a moderate probability that
the collected data is clean and  high quality.
Consequently, other prior distributions that place non-negligible probability at regions close to zero should be considered. Such distributions include the half-Cauchy distribution, uniform distribution, and truncated normal distribution. 

\subsubsection{Distributions for Mass Inflow Priors $q$}
The mass inflows $q$ to the network must be positive. Therefore, uniform and truncated normal distributions can be used. %
When there is little information about $q$, the upper bound can be set to a large value for both distributions. Alternatively, both the half-Cauchy and truncated normal distribution can be used without an upper bound, and set such that the shape of $p(q)$ is close to being flat.

\subsection{Posterior Sampling}

Once the prior and likelihood are established, the Bayesian inference problem can be solved as stated in \cref{e:Bayes}, updating our knowledge about our model parameters through the posterior distribution. 
Attempting to compute the posterior PDF would entail evaluating the denominator (model evidence) in Bayes' rule: $p(y)= \int p(y|\theta)p(\theta)\,d\theta$, a task that is generally intractable to perform even numerically except for very low (e.g., $<3$) dimensions of $\theta$.
Instead of computing the PDF, a major alternative strategy is to generate samples of $\theta$ from the posterior distribution. To that end, Markov Chain Monte Carlo (MCMC) algorithms~\cite{Gilks1996,Andrieu2003,Robert2004,Various2011} have become the predominant methods for computational Bayes in moderate $\theta$ dimensions (e.g., up to $\sim 100$), which iterates a Markov chain to generate samples that are consistent with the targeted posterior distribution. 
The more scalable MCMC methods include Hamiltonian Monte Carlo (HMC) based samplers such as the No-U-Turn (NUTs) algorithm, which explore the parameter space efficiently leveraging the posterior gradient and Hamiltonian energy principles \cite{betancourt2018conceptual}. However, HMC suffers from divergence in its time integration step when encountering neighborhoods of high posterior curvature \cite{betancourt2018conceptual, livingstone2018geometric}; this difficulty was indeed observed in our study when incorporating the data noise parameters $\sigma$ into the Bayesian inference. 
Therefore, we opt to use sequential Monte Carlo (SMC) \cite{doucet2001introduction} to sample the posterior, 
which is based on the idea of iteratively re-weighing the samples using on a tempered likelihood $[p(y|\theta)]^{\beta}$ at each stage where $\beta$ is a tempering parameter that gradually increases from 0 to 1.
We describe the SMC algorithm in SI Section 4.

\section{Case Study on the U.S. Steel Flow} 
\label{s:USGS}

The advances in the Bayesian inference approach to MFA discussed above are tested by mapping the U.S. annual flow of steel, where we take the MFA network structure from Zhu \textit{et al.}'s \cite{Yongxian} analysis of U.S. steel flows in 2014. 
The case study demonstrates rigorous prior development based on expert elicitation and inference of the MFA collected data noise as well as the MFA flow parameters. All the data and code used in this case study are available online (see the SI).

In this case study, the parameters requiring prior formulation are the allocation fractions ($\phi$), the external inputs ($q$), and the data noise standard deviation ($\sigma$) associated with data noise ($\epsilon$). 

\subsection{Constructing Priors for Allocation Fractions ($\phi$) and Input Flows ($q$)}

Since the elicitation of informative priors for all MFA variables of interest can be time prohibitive, we only elicit expert priors for the upstream allocation fractions ($\phi$) and external inputs ($q$), while using weakly informative priors elsewhere. Experts on U.S. steel flows were identified by conducting a literature search on steel flows and recycling and by contacting U.S. steel companies (e.g., U.S. Steel and Nucor). All the experts had more than 5 years of working or research experience in the steel industry. Eight experts agreed to an emailed request to take part in the study (see Table 2 in the SI). 

For prior construction, the experts independently completed surveys online that contained a total of 32 questions: 23 elicitation questions for allocation fractions associated with import, export, production and consumption of ferrous raw materials, and an extra 9 seeding questions whose actual observations are taken from USGS. 

\begin{figure}[htb]
	\centering
	\includegraphics[width=80mm, height= 45mm]{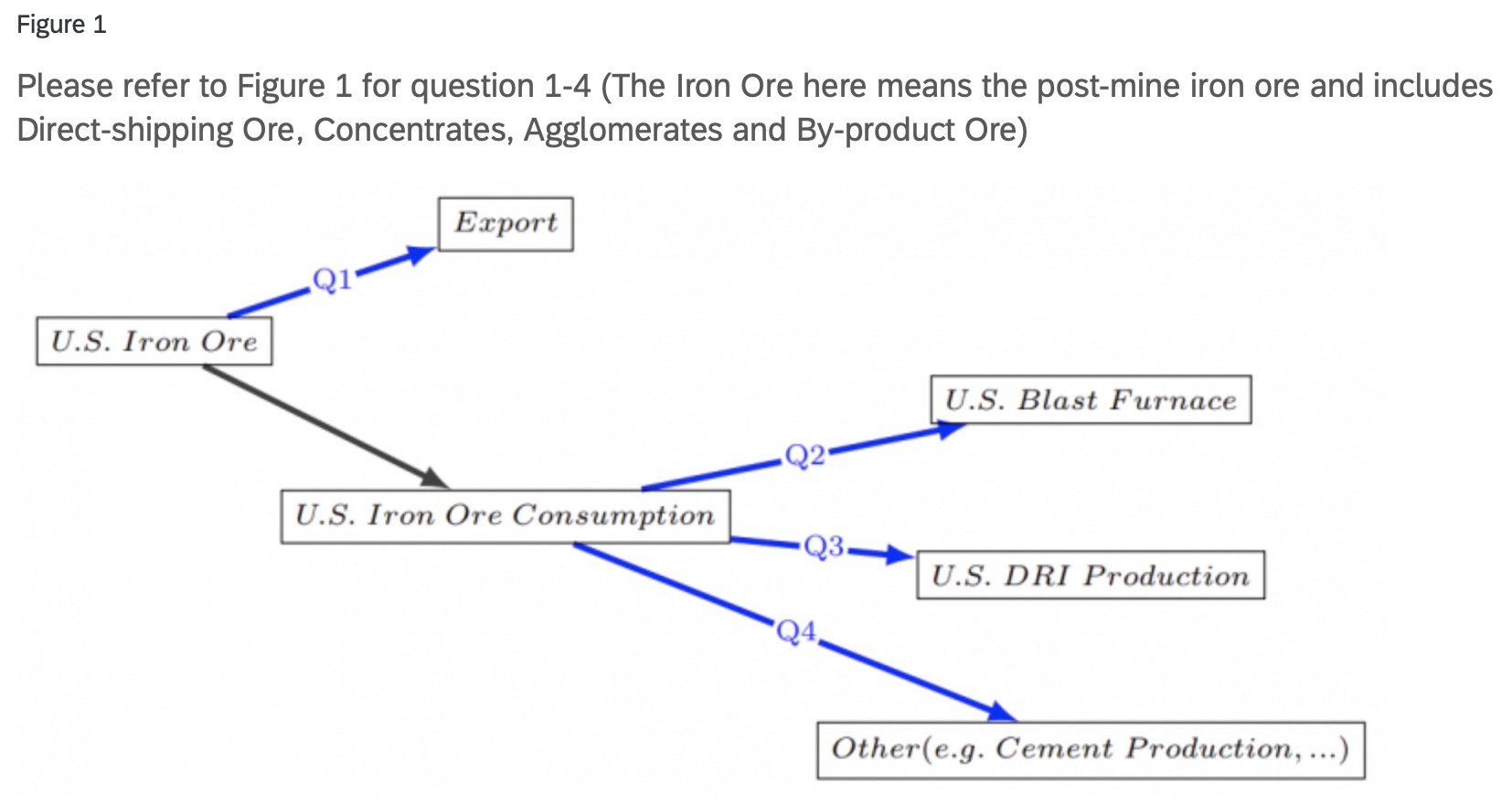}
	\includegraphics[width=80mm, height= 45mm]{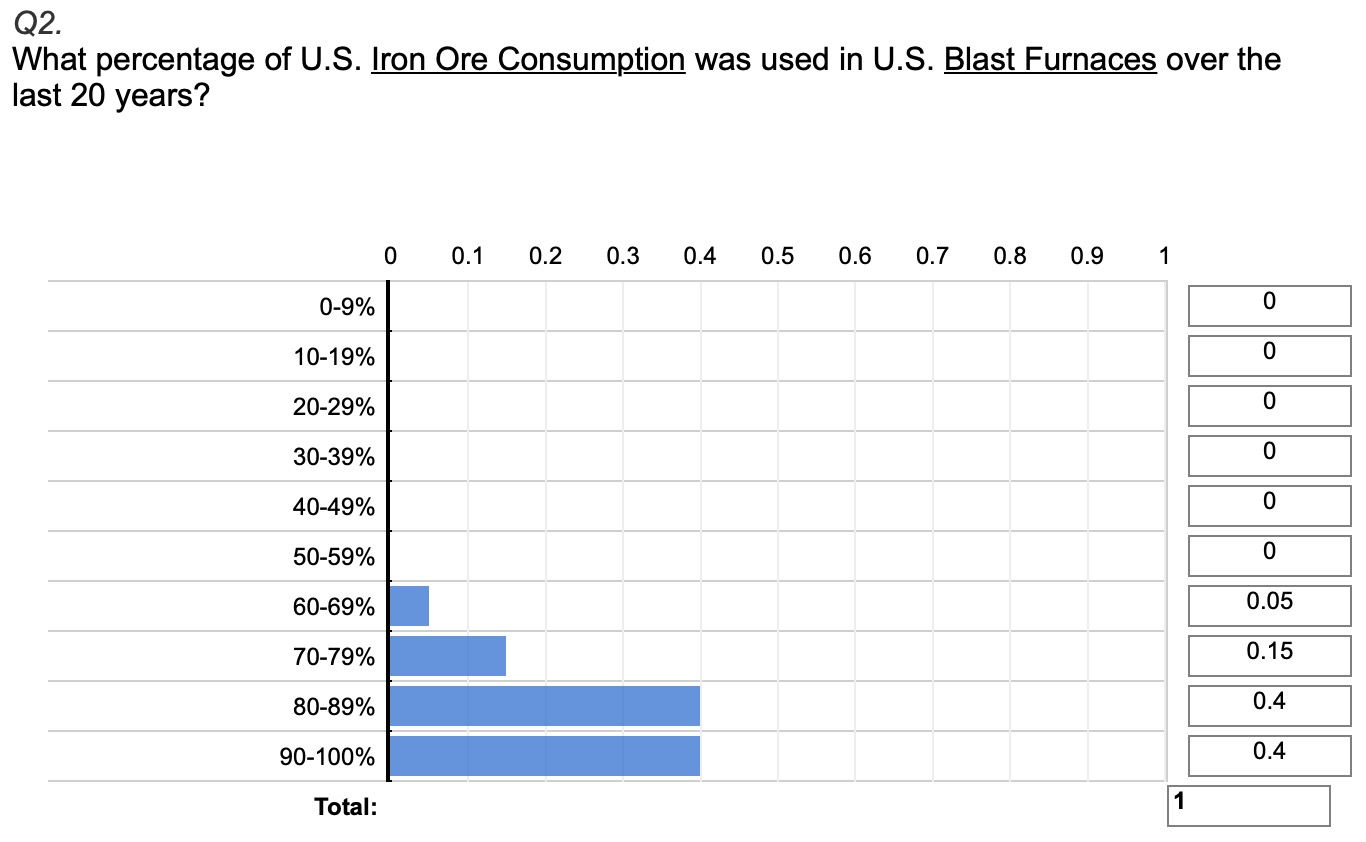}
	\includegraphics[width=170mm]{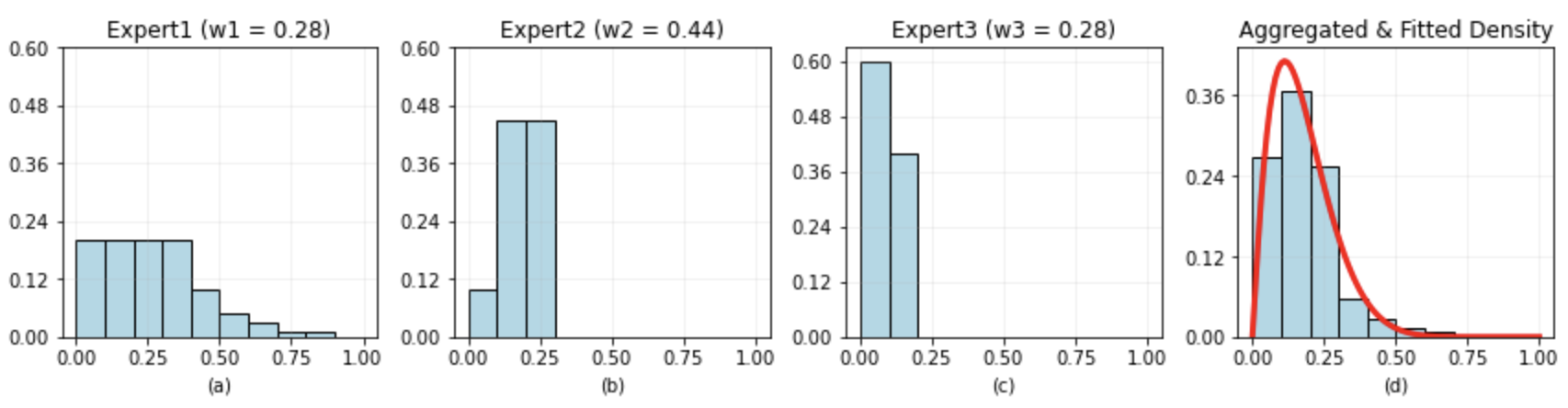}
    \caption{Prior elicitation required steel industry experts to complete an online Qualtrics Survey. The images above show an example question and expert response. The images below show an example calculation of an aggregated prior for the fraction of U.S. Iron Ore that is exported. Only 3 expert responses are shown for the sake of concision. Their weights have been normalized accordingly to sum to unity.}
  \label{f: ques}
\end{figure}

At least one author was present online to answer any questions during survey completion. It took the experts between 25 and 80 minutes each to complete the survey. The \textbf{Fixed Interval Method} was used to elicit the parameters. For allocation fractions $\phi$, the support $[0, 1]$ for $\phi$ was divided into 10 equal-width intervals and Dirichlet hyperparameters were fit to the elicited experts' histograms as described in \cref{ss:elicitation}. For the external inputs, the expert first specifies the lower and upper bound, with the interval then divided into 10 equal-width intervals. \Cref{f: ques} shows an illustrative example of a survey question for eliciting an allocation fraction. An expert could enter the probability value for each interval in the box or else drag the bar across for the given interval until the summation of the bars is 1, otherwise the expert cannot continue to the next question.

Linear pooling was used to aggregate the responses from the multiple experts into a single proper prior PDF for each MFA variable. First, weights were assigned to each of the experts based on their responses to 9 seeding questions using Cooke's method. A wide range of expert weighting values were obtained with the most informed expert having a weight of 0.299 and the least informed expert having a weight of less than 0.001.  \Cref{f: ques} shows how the response of multiple steel experts are combined to form a single aggregated prior PDF for an MFA variable.

\subsection{Collecting MFA Data Records and Constructing the Prior for Data Noise Standard Deviation ($\sigma$)}

After construction of the allocation fraction and external inflow priors, U.S. steel flow MFA data were collected. We focus here on 2012 data although any year from the previous 20 years could also have been chosen. The steel flow data were collected from the United States Geological Survey (USGS)  \cite{USGSIronOre12, USGSSteel12, USGSScrap12}, World Steel Association (WSA) \cite{WorldSteel12} and Zhu \textit{et al.} \cite{Yongxian}. A complete record of all the collected MFA data is provided in SI Section 5.

The collected MFA data are published without accompanying uncertainty information while data error is inevitable. Subsequently, the noise for each piece of collected data is modeled as an independent relative error (see \cref{e:relative_error}) that follows a Gaussian distribution with zero mean and a standard deviation $\sigma$. Expert elicitation could be used to derive an informed prior for $\sigma$; e.g., experts could be interviewed on the likely accuracy of USGS and WSA data. In this study, however, the prior on $\sigma$ is instead modeled as only weakly informative using a normal distribution truncated below zero and above 0.5 with hyperparameters set such that $\mathbb{P}(\sigma \leq 0.1)$ and $\mathbb{P}(\sigma \leq 0.3)$ are approximately 0.5 and 0.95 respectively, maintaining a reasonable probability that the data can be of high quality.

\subsection{Case Study Results: U.S. Steel flow in 2012}

\begin{sidewaysfigure}
  \centering
  \includegraphics[width=235mm]{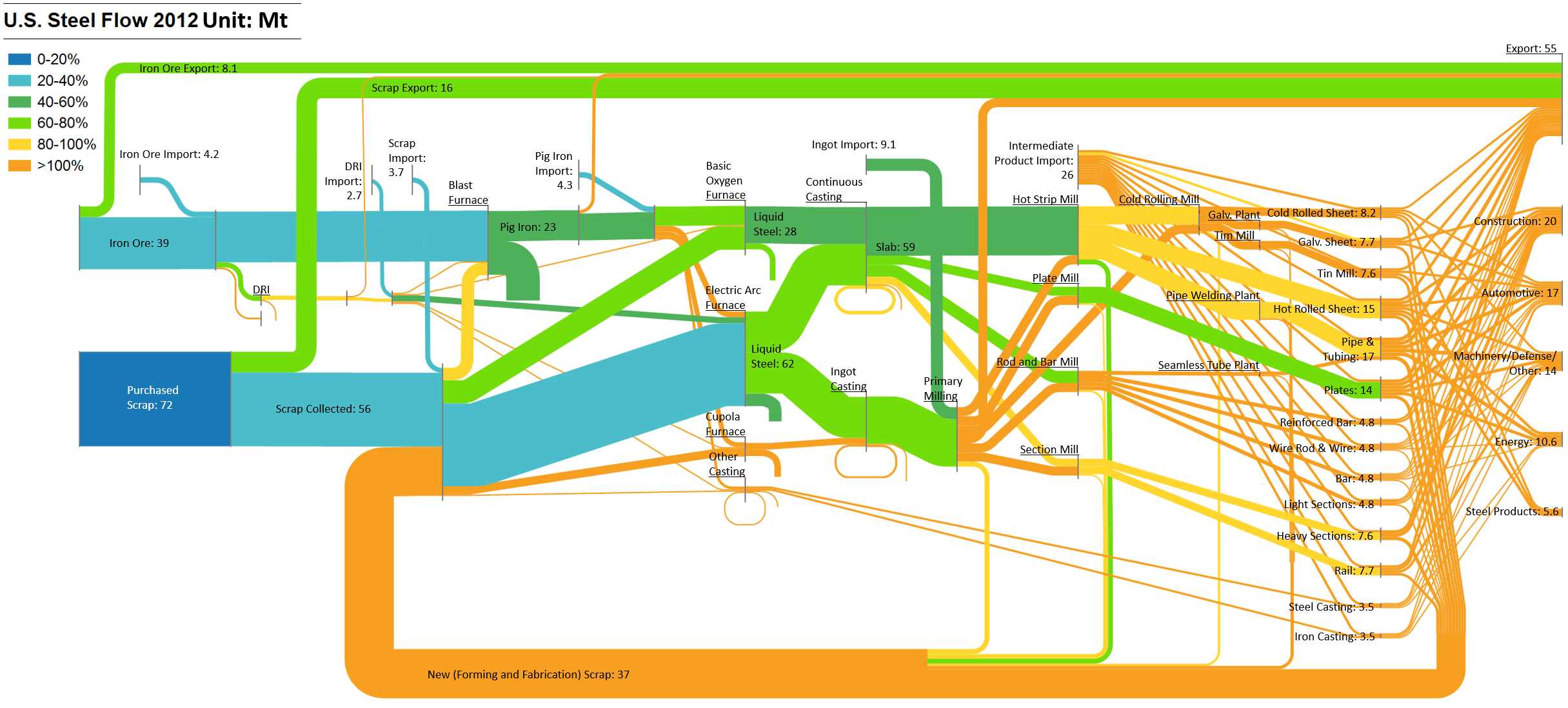}
  \caption{The prior for the U.S. Steel flow in 2012. All numbers on the flows refer to the mean of the prior mass flow in units of million metric tons (Mt). The uncertainty percentages refer to the flow standard deviation as a percentage of the flow mean. All mass flows refer to steel except for the iron ore flows that include the non-iron mass (e.g., oxygen and gangue).}
  \label{fig: Prior_2012}
\end{sidewaysfigure}

\begin{sidewaysfigure}
  \centering
  \includegraphics[width=235mm]{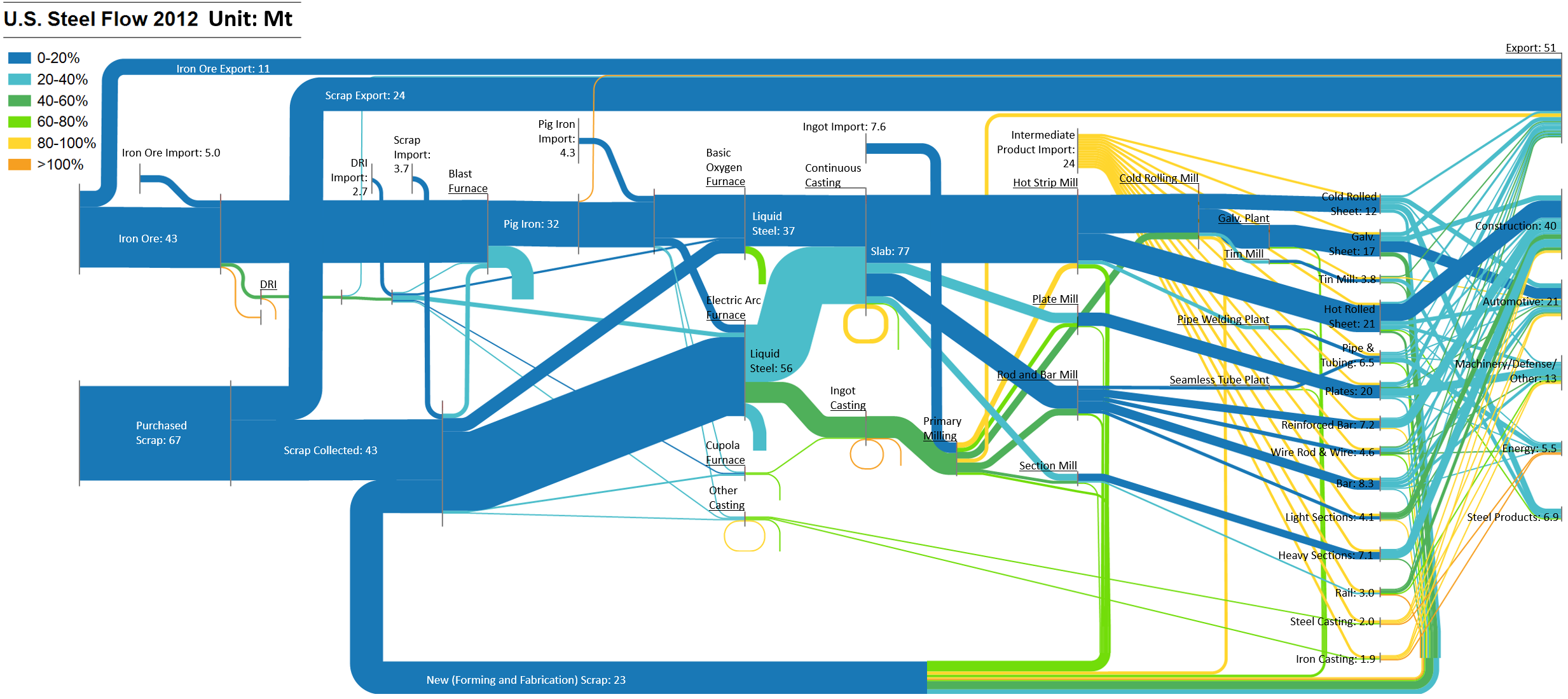}
  \caption{The posterior for the U.S. Steel flow in 2012. All numbers on the flows refer to the mean of the prior mass flow in units of million metric tons (Mt). The uncertainty percentages refer to the flow standard deviation as a percentage of the flow mean. All mass flows refer to steel except for the iron ore flows that include the non-iron mass (e.g., oxygen and gangue).}
  \label{fig: Posterior_2012}
\end{sidewaysfigure}

The Bayesian inference is implemented using SMC in PyMC3 with the code adapted from Lupton and Allwood \cite{lupton2018}. It takes approximately 17 hours to generate 10,000 samples
using an Intel(R) CoreTM i7-11800H CPU, 2.30 GHz. The prior and posterior results are shown in \cref{fig: Prior_2012,fig: Posterior_2012} respectively. The width of each line is proportional to the mean of the flow and the color indicates the uncertainty level, with a smaller relative uncertainty displayed in darker blue colors. 


\begin{figure}[htb]
   \centering
   \includegraphics[width=150mm]{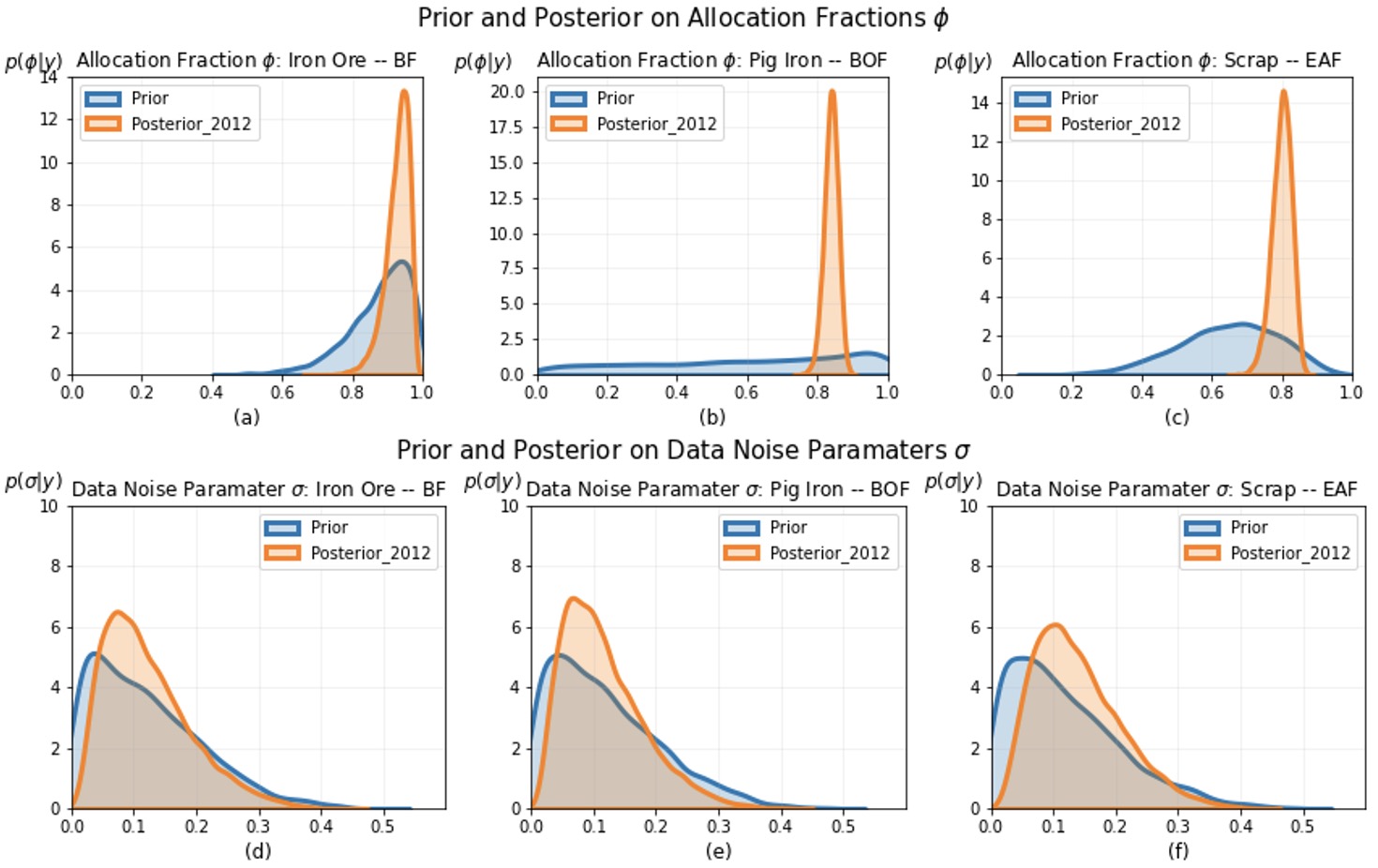}
   \caption{Examples of prior and posterior on Allocation Fractions ($\phi$) and Data Noise Parameters ($\sigma$) for Annual U.S. Steel Flow (2012).} 
  \label{f: Comparison_allocation_and_error}
\end{figure}

\Cref{f: Comparison_allocation_and_error} shows the prior and posterior distributions for $\phi$ and $\sigma$ associated with three prominent upstream flows. There is a significant uncertainty reduction for all $\phi$ shown in \cref{f: Comparison_allocation_and_error} (a)--(c). For example, despite a relatively flat prior for the fraction of (solid or liquid) pig iron consumed in the Basic Oxygen Furnace (BOF), even with only 1 year of data the posterior is able to reflect that pig iron is mainly consumed in the BOF. \Cref{f: Comparison_allocation_and_error} (d)--(f) presents the prior and posterior for $\sigma$. The posterior distribution for $\sigma$ on ``Iron Ore $\rightarrow$ BF'' is more concentrated 
than the prior, indicating that the data noise can be learned from data.
However, in comparison to $\phi$, there is a lower uncertainty reduction in $\sigma$. In an effort to enhance data noise learning, we explored the use of multiple years of data to prompt a greater reduction in uncertainty. 

\subsection{Enhanced learning using data collected from multiple years} 

One interpretation of the modest residual uncertainty for $\sigma$ shown in \cref{f: Comparison_allocation_and_error} is that there is still a low amount of information from data for inference if using data only from 2012. However, some of the $\phi$ are very likely to be similar across years.
In addition, the measurement noise $\epsilon$ in different years is also possibly a realization from a similar distribution. If the allocation fractions $\phi$ and data noise parameter $\sigma$ on regularly reported MFA data (e.g., the USGS data record on ``Iron ore $\rightarrow$ BF'') can be verified to be similar across a multi-year time period, then there is the potential to leverage multiple years' worth of MFA data to enhance the learning of the data noise. Therefore, we used Bayes factor analysis (see SI Section 3) to check and justify modeling the allocation fractions and data noise parameters as constant across five years worth of USGS and WSA data (2012--2016), allowing the inference to be rerun using these additional years' data.
 
\begin{figure}[h]
	\centering
	\includegraphics[width=150mm]{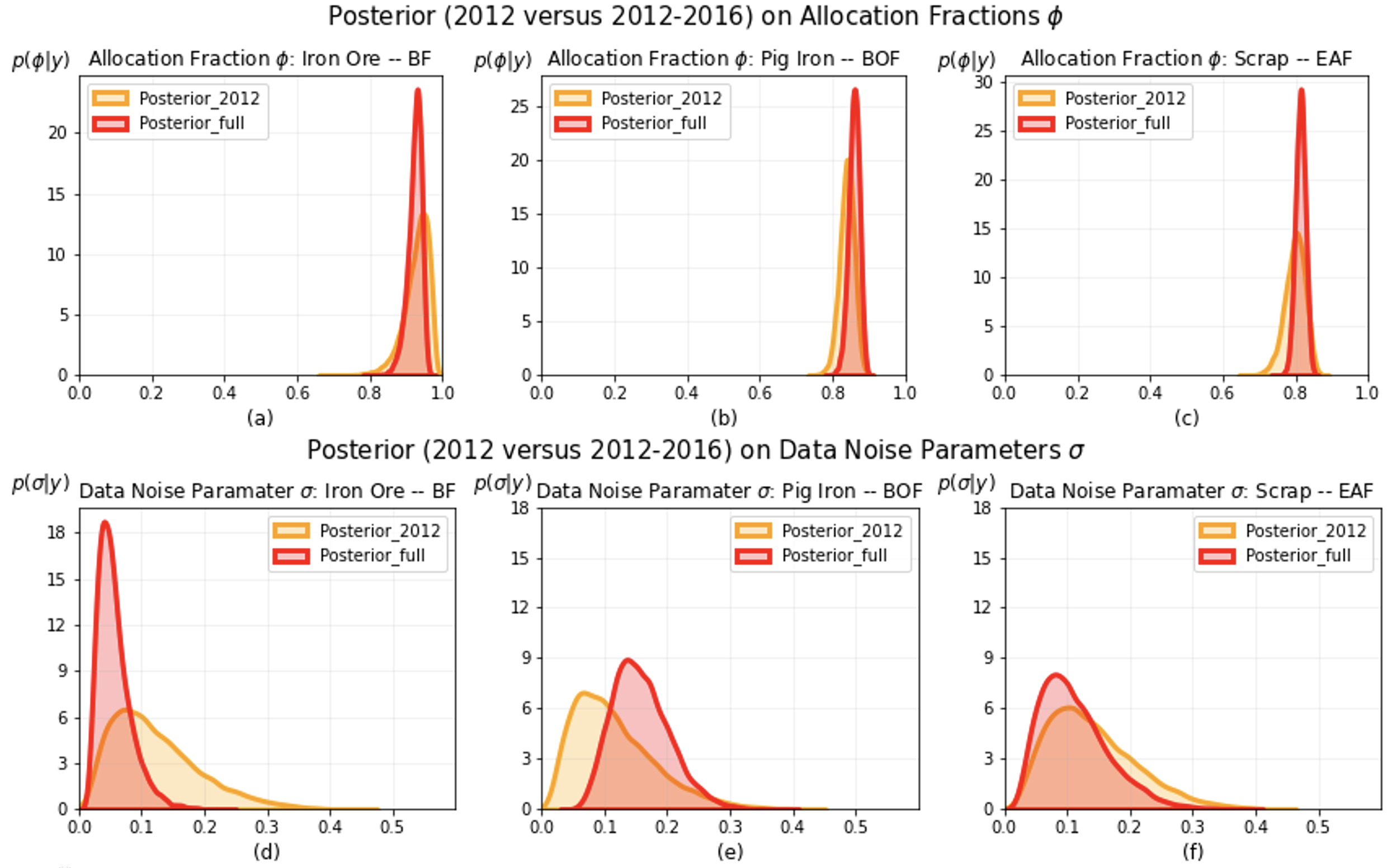}
  \caption{Examples of posteriors on Allocation Fractions ($\phi$) and Data Noise Parameters ($\sigma$) using data only from 2012 versus full data from 2012--2016.}
  \label{f: Comparison_allocation_and_error_5year}
\end{figure}

\Cref{f: Comparison_allocation_and_error_5year} shows the posteriors on $\phi$ and $\sigma$ when utilizing one years (2012) versus five years worth (2012--2016) of data. \Cref{f: Comparison_allocation_and_error_5year} shows a reduction in $\sigma$ uncertainty when leveraging these extra data. For ``Iron ore $\rightarrow$ BF'', utilizing 2012--2016 data reduces both the uncertainty on $\sigma$ and its mean value. On the other hand, the posterior on ``Pig Iron $\rightarrow$ BOF'' shows a reduced uncertainty for $\sigma$ but an increase in its mean. 
This could be because the data noise is indeed low in the ``Iron ore $\rightarrow$ BF'' data and high in the ``Pig Iron $\rightarrow$ BOF'' data. However, readers should note that the data noise calculated here reflects not only the collected MFA data quality but also any inadequacy in the modeling; e.g., from the MFA network structure used to the assumption of constant data noise errors across the five years.

\section{Discussion and Conclusions}
\label{s:discussion}

This paper has introduced, adapted, and demonstrated the use of expert elicitation techniques to define proper prior distributions for Bayesian inference in MFA, and illustrated how the noise level
in collected MFA data may itself be learned from data. Below, we discuss the lessons learned from the case study on U.S. steel flows, limitations of these approaches, and future work to advance the Bayesian approach in MFA.

\subsection{Using Expert Elicitation in MFA}
Expert elicitation allows the use of informed priors in MFA and will result in a quick reduction in parametric uncertainties when combined with collected data. In cases where no or negligible MFA data can be collected, expert elicitation provides a statistically robust method of estimating material flows. Expert elicitation may also reveal model-form mistakes in the MFA structure that were not apparent beforehand.  

A potential drawback is the time needed to find experts, develop well-posed elicitation questions, and process the responses. For the case study, the authors were  concerned that each interview might take more than one hour and that there would be significant confusion regarding our request to answer questions with histograms and PDFs rather than point values. However, most interviews were completed within 30 minutes and the authors only received approximately one survey clarification question per expert. Even experts who were unaccustomed to PDFs were comfortable completing the questions after reviewing the example question and solution we posted at the start of the survey. There was one case where the sum of upper bound elicited allocation fractions from a single source node was less than unity. While this does not prevent hyper-parameter fitting to a Dirichlet PDF it does suggest that either the expert made a mistake or that the expert thought that the model structure was incorrect, believing that there should be another destination node from that source node. A more sophisticated elicitation procedure could ask experts to evaluate the probability of candidate MFA model structures in addition to MFA parametric values. Overall, the Fixed Interval Method used in the case study was found to be an effective and quick method of eliciting allocation fraction and external inflow priors. 

In the case study, a single weight was assigned to each expert. A potential problem is that the expert might not be uniformly informed across the domain of interest; e.g., an expert might be an authority on steel recycling but relatively uninformed on primary steelmaking. In the case study, it was possibly a case of non-uniform expertise that affected the derivation of the aggregated prior for the allocation fraction from (liquid and solid) pig iron to the BOF.
The expert who received the highest weighting gave an apparently erroneous answer to the elicitation question on ``Pig Iron $\rightarrow$ BOF'', responding that less than 50\% of the iron flows into the BOF. The high weighting given to this expert meant that the aggregated prior for Pig Iron to BOF was relatively uninformed (flat) despite the other seven experts responding that the majority of pig iron flows into the BOF (see \cref{f: Comparison_allocation_and_error}). One option to avoid such problems would be to assign multiple weights to each expert corresponding to different areas of expertise; however, that would require the development and asking of more seeding questions tailored to those different areas of expertise.  

A source of confusion for one of the experts was whether the survey was asking for uncertainty on a given variable 
(e.g., $\theta = \frac{\text{Amount of Iron Ore Exported}}{\text{Amount of Iron Ore Produced}}$) 
or its variability
over space and time; e.g., the histogram of the $\frac{\text{Amount of Iron Ore Exported }}{\text{Amount of Iron Ore Produced}}$ for each month in a year or across regions. The survey was eliciting the former and we found it useful to clarify that we were seeking uncertainty rather than variability by demonstrating a ``toy'' example and also using equations to define the variable of interest. 

\subsection{Learning the MFA Variables and Data Noise Parameters}
The uncertainty reduction in the U.S. steel mass flows between the prior and posterior is shown by the increase in dark blue colors between  \cref{fig: Prior_2012,fig: Posterior_2012}. The ability to quantify and reduce these uncertainties in a principled manner using Bayesian inference is appealing as it can lead to more informed decision- and policy-making. For example, the benefits of deploying a new, more energy efficient mill technology can be calculated with greater confidence for the cold rolling mill (calculated to have processed an expected 32.3 Mt in 2012 with a standard deviation of 3.27 Mt) compared to the primary mill  (calculated to have processed a smaller expected value of 23.1 Mt in 2012 but with a larger standard deviation of 8.69 Mt).  

A new MFA approach explored in this article has been to incorporate data noise parameters as random variables. The allocation fractions and the data noise parameters are then learned simultaneously with collected MFA data. Intuitively it is harder to achieve great uncertainty reductions using this method because the same amount of data is used to provide information on more parameters, even though it is a more honest representation if we do not know the true data noise.
To combat this, we used Bayes factor analysis to verify the existence of time-invariant data noise parameters, allowing us to incorporate multiple years worth of data for greater uncertainty reductions.

As observed elsewhere \cite{lupton2018}, a potential drawback of the Bayesian approach to MFA is the computational cost of the Monte Carlo-based algorithms. Modeling the data noise parameters as random variables significantly increased the stochastic dimension of the problem and in turn computational cost, from 3 hours/run if the data noise parameters are prescribed as constants, to 17 hours/run when the data noise parameters are modeled as random variables. Therefore, there is a trade-off between the amount of bias and the computational cost. However, the computational speed can be increased by, for example, using multi-core processors to run algorithms that can be parallelized or applying approximate Bayesian techniques such as variational inference~\cite{Blei2017}.

\subsection{Future Work}

The Bayesian approach to MFA provides a mathematically principled procedure to incorporate expert knowledge alongside sparse, noisy, and often incomplete data records: a data-informed model learning approach. We plan to investigate how this model-and-data relationship can be leveraged to create intelligent data acquisition strategies for seeking out the most informative data that can tell us what the MFA structure should look like, and what its parameters values are. These remain important challenges in MFA and can be approached by combining the use of Bayesian inference for uncertainty quantification early in the MFA exercise with the principles of Bayesian experimental design~\cite{Chaloner1995,Muller2005}.

\section{Acknowledgements} 
The authors would like to thank all the steel experts interviewed for this study. This material is based upon work supported by the National Science Foundation under Grant No. \#2040013.

 \section{Supporting Information} 
The Supporting Information (SI) includes data and literature reviews helpful to understanding the main article as well as links to elicitation surveys and Python Scripts that allow other MFA practitioners to use expert elicitation and data noise learning techniques.

\clearpage
\bibliography{ms}
\bibliographystyle{unsrturl}

\end{document}


\maketitle

This Supporting Information (SI) document includes data and literature reviews helpful to understanding the main article as well as links to elicitation surveys and Python Scripts that allow other MFA practitioners to use expert elicitation and data noise learning techniques.

Please go to this link (\url{http://remade.engin.umich.edu/MFA_NSF.htm}) for downloads of the following: 

\begin{itemize}
    \item Presentation of the underlying data used to construct the Sankey diagrams in the main paper (Figures 5 and 6) in a numerical, tabular format
    
    \item An example Google Qualtrics survey for eliciting allocation fractions from experts
    
    \item An Excel file for calculating expert weights based on seeding variable question responses
    
    \item A Python script for
fitting prior PDFs to the aggregated and weighted histograms from experts

\item A Python script for performing Bayesian inference (adapted from Lupton and Allwood \cite{lupton2018}) using the fitted prior PDFs and USGS and WSA collected MFA data

\item A Python script for performing Bayes factor estimation to select best performing model assumptions

\end{itemize}

\section{Variable versus Fixed Interval Elicitation Methods}

\Cref{table:univariate} summarizes the implementation of some common approaches for eliciting a univariate $\theta$ from an individual expert.  

One of the most widely used Variable Interval Methods is the Bisection method, which asks for medians of different intervals and is straightforward for experts to understand.
However, some empirical studies reported overconfidence with the Bisection in comparison to the Tertile method \cite{OHagan2006a}, which operates similarly to the Bisection method but with more refined intervals. 
For the Fixed Interval Method, there are various options to define the intervals that are presented to experts to obtain their probability estimates. In \cite{SHELF, OHagan1998}, intervals are chosen with bins concentrated around the median or mode of $\theta$, and focus on the elicitation of regions where the bulk of probability lies. The probabilities of $\theta$ within different intervals are asked in specific order to avoid tendencies of anchor assessments, which describes the phenomenon that an expert may adjust their estimates for later questions based on their own earlier responses. 
The trail roulette method is a Fixed Interval Method which does not ask for specific probabilities \cite{gore1987biostatistics}. Instead,  it requires experts to distribute a number of chips among the bins of a histogram, and the proportion of total chips for a bin then represents the expert's probability of the parameter lying within that bin. The procedure may be simplified by increasing the width of bins or decreasing the number of chips although at the expense of losing resolution in representing the distribution.
 
\begin{table}
 \centering
 \begin{tabular}{|c|c|c|c|}
 \hline
 {\textbf{Type}} & {\textbf{Examples}} & {\textbf{Expect Response on $\theta$}} & {\textbf{Comments}} 
\\
 \hline\hline
 \multirow{4}{*}{\makecell{Variable Interval \\ Method}} &   
  Bisection Method \cite{raiffa1968} &  
 \makecell{lower and upper bound, \\  quartiles of $\theta$}  & 
 \makecell{Straightforward \\ to understand}
 \\\cline{2-4}
 & Tertile Method \cite{Garthwaite2000} &  \makecell{lower and upper bound, \\  tertiles of $\theta$}  & { \makecell{Less overconfidence \\
 than Bisection}}  \\
 \hline
 \multirow{8}{*}{\makecell{Fixed Interval \\ Method}} & \makecell{Option 1: SHELF \\ package \cite{SHELF}} & \makecell{ lower  ($L$) and upper ($U$) \\ bounds, median ($M$) of $\theta$, and \\ probability of $\theta$ in \\ $(L, \frac{2M+L}{3})$, $(\frac{2M+U}{3},U)$} &  \multirow{6}{*} {\makecell{More accurate \\ elicitation in region \\ where the bulk of\\ probability lies}} \\
 \cline{2-3}
 & \makecell{Option 2: \cite{OHagan1998}} & \makecell{lower  ($L$) and ($U$) upper  \\ bounds, mode ($R$) of $\theta$, and \\ probability of $\theta$ in $(L,R)$, \\ $(L, \frac{L+R}{2})$, $(\frac{R+U}{2}, U$, \\ $(L, \frac{L+3R}{4})$, $(\frac{3R+U}{4}, U)$} &
 \\\cline{2-4}
 &  \makecell{Trail Roulette Method \cite{gore1987biostatistics}}& \makecell{Allocate chips into \\ equal-width bins} & \makecell{Not  asking for \\specific probabilities } \\
 \hline
 \end{tabular} 
 \caption{Method for elicitation of a univariate $\theta$.}
 \label{table:univariate}
 \end{table}

There are a few experiments comparing Variable Interval Methods and Fixed Interval Methods but without conclusive findings for one being better. Murphy and Winkler \cite{MurphyandWinkler1974} invited 4 experienced weather forecasters to make credible interval forecasts for temperature in Denver. Two of them forecasted based on the bisection method and estimated the 12.5, 25, 37.5, 50, 62.5, 75, 87.5th percentiles, while the other two used the Fixed Interval Method and gave probabilities of $\pm5^{\circ}$C and  $\pm9^{\circ}$C intervals around a pre-determined median. In that study, results from the bisection method were much closer to the {observed} relative frequencies than the fixed interval forecasts. 
Separately, Abbas \textit{et al.} \cite{Abbas08} conducted a comparison experiment to assess the  temperature in Palo Alto with 72 students. 
The Variable Interval Method involved estimating the 5, 25, 50, 75 and 95th percentiles, and the Fixed Interval Method entailed intervals with boundary points at the 5\%, 25\%, 50\%, 75\% and 95\% of a total range specified by the judges. The order of questions for both methods were randomizes. When comparing with 345 historical data points, results from the Fixed Interval Method achieved a better fit, contradicting the findings from Murphy and Winkler~\cite{MurphyandWinkler1974}. Moreover, the Fix Interval Method was reported to induce faster responses,
and 50 participants indicated preference for the Fixed Interval Method while the other 22 either preferred the Variable Interval Method or showed no preference. 

\section{Elicitation from Multiple Experts}

\subsection{Inconsistency in the Linear and Logarithmic Opinion Pooling Methods} 

Both the linear and logarithmic opinion pooling methods suffer some form of inconsistency. Linear pooling is not externally Bayesian \cite{OHagan2006a}. That is, updating each expert's prior with data and aggregating their posteriors into single one would yield a different result than first combining the experts' priors 
and then updating the aggregated prior to the posterior. While the logarithmic pooling is externally Bayesian, it does not satisfy coherent marginalization. To satisfy coherent marginalization, for an event ``$C$'' equal to ``event $A$ or $B$ where $A$ and $B$ are mutually exclusive'' (and therefore $\PP(C) = \PP(A) + \PP(B)$), one can either compute $\PP(C)$ from each expert and then pool the result, or first pool $\PP(A)$ and $\PP(B)$ separately and then sum them up. Logarithmic pooling does not satisfy this marginalization property while linear pooling does \cite{McConway1981}. O'Hagan \textit{et al.} \cite{OHagan2006a} further point out that no mathematical formula can be simultaneously  externally Bayesian and abide coherent marginalization, and Lindley \cite{lindley1985} and Genest and Zidek \cite{genest1986combining} criticised the evaluation of these fundamental inconsistencies, arguing that the aggregated distribution does not represent the belief of a single individual and therefore need not to behave as one would expect an individual's probability distribution. 

\subsection{Performance of Cooke's Method} 

To assess the performance of Cooke's method, Cooke and Goossens \cite{Cooke2000} conducted a psychometric experiment comparing ``experienced experts'' (teachers at a technical training institution) versus ``inexperienced experts'' (students at the institution). On technical items, the experienced experts performed significantly better on both the calibration and information scores. On general knowledge items, there was no significant difference between the teachers and students. Therefore, Cooke's method appears effective at reflecting the teachers expertise. 

Cooke and Goossens \cite{Cooke08} also conducted studies comparing (i) Cooke's method {for assigning expert weights}, (ii) simple averaging (uniform weights), and (iii) putting all weight on the best expert only {as revealed using Cooke's method} (i.e. one with the highest original weight). 
In their work, 45 separate experiments were carried out. 
In each experiment, every seeding question's histograms from multiple experts were combined into a single histogram using the three aforementioned weighting methods. Under each method, its combined histograms were then used to compute the new information and calibration scores (in the same manner as that in Cooke's method), and the product of these two scores is used as the assessment metric for comparison. 
Cooke's method achieved the highest metric among the three weighting approaches in {27} of the 45 cases, and it scored higher than the uniform weighting approach in {44} of the 45 cases. However, 
by using an assessment metric that is the same weighting formula from Cooke's method (i.e. product of information and calibration scores), it may have automatically favored the competition towards Cooke's method. Furthermore, the assessment metric was calculated on the same seeding questions as those used 
to determine the expert weights,
and thus may 
not reflect the ability for the weighting approaches to generalize to priors for other quantities not covered in the seeding questions.

\subsection{Experts for Prior Elicitation in the Case Study}

Details of experts interviewed for prior elicitation in the case study of this paper are shown in \cref{table: Experts in elicitation}.
\begin{table}[h]
\centering
\begin{tabular}{|c| c c|} 
\hline
 \textbf{Expert} & \textbf{Affiliation} &  \textbf{Area of Expertise} \\
\hline\hline
 Expert 1 & University of Michigan & Sustainability in metals processing  \\
 \hline
 Expert 2 & \makecell{United States \\Geological Survey} &  Statistics for ferroalloys\\ 
 \hline
 Expert 3 & \makecell{United States \\Steel Corporation}
 & \makecell{Advanced materials \\ and manufacturing}  \\ 
 \hline
 Expert 4 & \makecell{United States \\Steel Corporation}
 &  \makecell{Advanced materials \\ and manufacturing}  \\ 
 \hline
 Expert 5 & AK Steel Corporation & \makecell{Ferroalloys, superalloys,\\ stainless steel, and advanced \\high strength steel}\\ 
 \hline
 Expert 6 & \makecell{Continuous Improvement \\ Experts}
 &  \makecell{Steelmaking and\\ refining technology} \\ 
 \hline
 Expert 7 & University of Cambridge  & \makecell{Energy systems, resource \\efficiency, demand reduction} \\ 
 \hline
 Expert 8 & Nucor Corporation & Automotive market and product development \\
 \hline
\end{tabular}  
\caption{Steel flow experts interviewed for prior elicitation.}
\label{table: Experts in elicitation}
\end{table}

\subsection{Special Discussion on Seeding Variable}

An interesting consideration for the Cooke's method that has not been discussed in existing literature is regarding the nature of the elicitation distribution.
In particular, there is inconsistency to
Cooke's method if the reference answer value to the seeding question is the true value to the quantity being asked.
For example, suppose a seeding question asks ``what is the value of $\pi$ up to tenth digit?'' and suppose the reference answer is the exact value 3.141592653, then a perfect expert who perfectly knows $\pi$'s digit then would provide a histogram that is centered around this value and with a minimally required uncertainty (e.g., due to the minimum bin width of histograms in the survey). If the expert answers all seeding questions optimally in this manner, the empirical interquantile interval distribution (the blue distribution in the middle column of Fig. 3 (b) in the main paper) would be concentrated near the median, and tail intervals (e.g., $p_1$ and $p_4$) would be zero. Hence, it would in fact be quite different from the ``ideal'' interquantile interval distribution suggested by Cooke (the red dotted distribution in the middle column of Fig. 3(b) in the main paper). This perfect expert would then obtain a penalized, sub-optimal calibration score. 
In contrast, a less informative expert who provides more uncertain distributions but able to achieving, e.g., $\{p_1, p_2, p_3, p_4\} = \{0.05, 0.45, 0.45, 0.05\}$ following our setup in the main paper, would have a higher calibration score. These outcomes conflict with our intuition and expectations. 

The root of this inconsistency lies in that Cooke's method, by placing the reference answer values in the interquantile intervals provided by the expert, is effectively checking whether the reference answer values are drawn from the expert's distributions. This notion could be consistently incorporated if the seeding question asks the expert to provide an observation distribution from which the reference answer is obtained. The question then must also supply the context about the source of the reference answer value; e.g., if it were acquired by USGS or WSA, etc. However, while more statistically precise, such questioning would be highly complex and difficult to explain and understand by a human expert.

\section{Bayesian Factor for Justifying Using Data from Multiple Years for Inference} 

The Bayesian framework can also be used to compare different models and their assumptions by computing \emph{Bayes factors}. In this work, we will use Bayes factors to compare among models where the allocation fractions and noise parameters are assumed to be time-invariant (i.e. same across multiple years) and when they are time-dependent. If the former hypothesis is supported by the data, we can potentially use multiple years' MFA data together to inform these model parameters. 

We begin by restating the Bayes' rule from Eqn. (5) of the main paper that describes the update of uncertainty for the model parameters, but now making it explicit that the expression is for a given model $M_i$ (and its associated assumptions, structure, and parameterization):
\begin{align}
    p(\theta_i|y,M_i) = \frac{p(y|\theta_i,M_i)p(\theta_i|M_i)}{p(y|M_i)}
    \label{e:Bayes_M}
\end{align}
with $\theta_i$ indicating these are the parameters for model $M_i$ (a different model $M_j$ may be associated with different parameters $\theta_j$).
Furthermore, we can also write the Bayes' rule for the \emph{model}: 
\begin{align}
    \PP(M_i|y) = \frac{p(y|M_i)\PP(M_i)}{p(y)}
    \label{e:Bayes_for_models}
\end{align}
where $\PP(M_i)$ represents the prior belief that model $M_i$ is the data-generating model, $\PP(M_i|y)$ is the updated posterior belief given data $y$, and $p(y|M_i) = \int p(y|\theta_i, M_i)p(\theta_i|M_i)\, d\theta_i$ is the \emph{Bayesian model evidence}. This model evidence term is also precisely the denominator term in the parameter-Bayes' rule in \cref{e:Bayes_M}. 
Then, we can compare between two models their probabilities of being the data-generating model given observed data:
\begin{align}
    \frac{\PP(M_1|y)} {\PP(M_2|y) } = \frac{p(y|M_1)}{p(y|M_2)} \frac{\PP(M_1)}{\PP(M_2)},
\end{align}
which if assuming a uniform model prior ($\PP(M_i) = \PP(M_j)$) simplifies to the ratio of model evidence:
\begin{align}
    \textrm{BF}(M_1:M_2) = \frac{p(y|M_1)}{p(y|M_2)},
\end{align}
which is known as the \emph{Bayes factor}.
A Bayes factor here greater (smaller) than 1 represents an increase (decrease) of support in favor of model 1 versus model 2 given the observed data. Robert \textit{et al.} \cite{Jeffrey} discusses a scale originally proposed by Jeffreys to interpret Bayes factors according to the strength of evidence in favor of one model, for example, a Bayes factor $>100$ is ``extreme evidence for $M_1$'', and 30--100 is ``very strong evidence for $M_1$''. However, this scale and their ranges provide a guideline and should not be regarded as an absolute rule.

For our paper's case study, we compare four candidate models: (1) all allocation fractions $\phi$ and data noise parameters $\sigma$ are invariant between 2012--2016; (2) all $\phi$ are invariant but $\sigma$ are different each year between 2012--2016; (3) all $\phi$ are different each year but parameters $\sigma$ are invariant between 2012--2016; and (4) all $\phi$ and $\sigma$ are different each year between 2012--2016. Uniform prior is adopted for these model hypotheses (i.e. $\PP(M_i) = 1/4$). Computing the Bayes factors then entails estimating the model evidence, which can be obtained through a Monte Carlo estimator:
\begin{align}
    p(y|M_i) = \int p(y|\theta_i, M_i)\, p(\theta_i|M_i)\, d\theta_i \approx \frac{1}{N}\sum_{j=1}^N p(y|\theta_i^{(j)}, M_i)
    \label{e: BayesFactor}
\end{align}
where $\theta_i^{(j)}$ are samples drawn from the prior $p(\theta_i|M_i)$, and $p(y|\theta_i^{(j)}, M_i)$ is the same likelihood PDF described in Sec. 2.3 of the main paper which we know how to evaluate.

\begin{table}
 \centering
 \begin{tabular}{|c|c|c|c|}
 \hline
   & Min & Mean & Max \\
 \hline
 {BF($M_1:M_2$)} & $5.28 \times 10^{13,973}$ & $4.27 \times 10^{49,667}$ & $9.03 \times 10^{84,114}$ \\
 \hline
 {BF($M_1:M_3$)} & $2.23 \times 10^{350,062}$ & $2.07 \times 10^{1,271,999}$ & $7.49 \times 10^{2,909,082}$ \\
 \hline
 {BF($M_1:M_4$)} & $7.37 \times 10^{664,549}$ & $5.17 \times 10^{2,061,424}$ & $4.90 \times 10^{4,127,542}$ \\
 \hline
 {BF($M_2:M_3$)} & $8.98 \times 10^{315,700}$ & $4.85 \times 10^{1,222,331}$ & $5.06 \times 10^{2,826,606}$ \\
 \hline
 {BF($M_2:M_4$)} & $1.40 \times 10^{650,576}$ & $1.21 \times 10^{2,011,757}$ & $1.93 \times 10^{4,072,514}$ \\
 \hline
 {BF($M_3:M_4$)} & $1.26 \times 10^{-874,603}$ & $2.50 \times 10^{789,425}$ & $1.20 \times 10^{2,277,781}$ \\
 \hline
 \end{tabular} 
  \caption{Bayes factors to compare among models where the allocation fractions and noise parameters are assumed to be either  time-invariant or time-dependent across the U.S. steel sector (2012--2016). Model 1 is that all allocation fractions $\phi$ and data noise parameters $\sigma$ are time-invariant; Model 2 is that all $\phi$ are invariant but $\sigma$ are different each year; Model 3 is that all $\phi$ are different each year but parameters $\sigma$ are invariant; Model 4 is that all $\phi$ and $\sigma$ are different each year. The minimum, mean, and maximum values are from repeating 30 trials of $10^5$-sample Monte Carlo estimate.  A $BF(M_A : M_B)$ greater (smaller) than 1 represents an increase (decrease) of support in favor of model $A$ versus model $B$ given the observed data. In this case, model 1 receives overwhelming support in comparison to all other models.}
 \label{table:BF}
\end{table}

\Cref{table:BF} presents the Bayes factors for all model pairs from the four models considered, computed using 
{$10^5$} Monte Carlo samples implemented with the log-sum-exp technique that alleviates numerical underflow. 
The Bayes factor estimates are also repeated for {30} trials, and the minimum and maximum values are also reported in \cref{table:BF}. The extremely large Bayes factors observed suggest overwhelming support for model 1 where all $\phi$ and $\sigma$ are invariant between 2012--2016, and then in descending order followed by models 2, 3, and 4. 
These results justify the use of multiple years' data from 2012--2016 to learn $\phi$ and $\sigma$ as done in Sec. 3.4 of the main paper.

\section{Sequential Monte Carlo}

SMC evolves a population of samples from the prior to the posterior. To achieve this, it employs an auxiliary temperature parameter $\beta$ which increases from 0 towards 1 over multiple stages, and the tempered posterior $p(\theta|y)_{\beta} \propto p(\theta)p(y|\theta)^{\beta}$ is the prior when $\beta = 0$ and becomes the posterior when $\beta = 1$. At stage $t$, we start with a population of $n$ samples from the current tempered posterior. The importance weight of each sample is then computed as the ratio of the tempered likelihoods between stage $t+1$ and $t$. The population is then re-sampled based on these weights, and each sample is also perturbed with a small number of steps in a Metropolis-Hastings Markov chain. This procedure is iterated until $\beta$ reaches 1. The SMC algorithm is summarized in \cref{alg:smc}. For this work, the SMC algorithm is implemented with PyMC3.

\begin{algorithm}[htb]
\caption{SMC algorithm.}\label{alg:smc}
\hspace*{\algorithmicindent} \textbf{Input:} \;\; Prior $p(\theta)$, likelihood $p(y|\theta)$, population sample size $N$, number of Metropolis-Hastings steps $n_{steps}$, effective sample size threshold $N_t$.  \\
 \hspace*{\algorithmicindent} \textbf{Output:}\; Samples from $p(\theta|y)$.
\begin{algorithmic}[1]
\STATE \textbf{Initialize:}
$\beta = 0$, $n = 0$\;, sample $N$ particles from $p(\theta)$.
\WHILE{$\beta < 1$}
 \STATE $n = n + 1$, increase $\beta$ according to tempering schedule.
 \STATE Reweigh particles by $\tilde{w}_n^i \propto \frac{p^{\beta}_n(y|\theta)}{p^{\beta}_{n-1}(y|\theta)}$  and normalize such that $\sum_{i=1}^N\tilde{w}_n^i = 1$.
 \STATE Estimate Effective Sample Size ($\widehat{ESS}) = \frac{N}{\frac{1}{N}\sum_{i=1}^N(\tilde{w}_n^i)^2}$
 \IF{$\widehat{ESS} < N_t$}
 \STATE Re-sample $N$ particles based on empirical distribution using $\tilde{w}_n^i$.
 \ENDIF
\STATE Perturb each particle for $n_{steps}$ steps of Metropolis-Hasting Markov chain.
\ENDWHILE
\end{algorithmic}
\end{algorithm}

\section{Case Study: U.S. Steel Flow MFA Collected Data}
\begin{center}
\begin{longtable}{  m{17em} m{3.5cm} m{3cm} m{2cm}  } 
  \hline
  \textbf{Description} & \textbf{Type} &\textbf{Value (Mt)} & \textbf{Source} \\ 
  \hline\hline
  Import to Iron Ore Consumption & External Input &  5.16 & 1 \\ 
  \hline
  Iron Ore Production & External Input & 54.7 & 1 \\ 
  \hline
  Iron Ore Production to Export & Flow & 11.2  & 1 \\ 
  \hline
  Iron Ore Consumption to Blast Furnace &  Flow  & 46.3  & 1 \\ 
  \hline
  Blast Furnace to Pig Iron &  Flow  & 32.1  & 3 \\
   \hline
  Import to DRI Consumption & External Input & 2.47 & 2 \\ 
  \hline
  DRI to Export &  Flow  &  0.01 & 2 \\ 
  \hline
  DRI Consumption to Blast Furnace &  Flow  & 0.049  & 2 \\ 
  \hline
  DRI Consumption to Basic Oxygen Furnace &  Flow   &  1.91 & 2 \\ 
  \hline
  DRI Consumption to Electric Arc Furnace &  Flow  &   1.62 &  2\\ 
  \hline
  DRI Consumption to Cupola Furnace &  Flow  & 0.01  & 2 \\ 
  \hline
  DRI Consumption to Other &  Flow  & 0.01  & 2 \\ 
  \hline
  Import to Pig Iron Consumption & External Input &  4.27 & 2 \\
  \hline
  Pig Iron to Export &  Flow  & 0.021  & 2\\ 
  \hline
  Pig Iron to Basic Oxygen Furnace &  Flow  &  31.5  & 2 \\ 
  \hline
  Pig Iron to Electric Arc Furnace &  Flow  &  5.79 & 2 \\ 
  \hline
  Pig Iron to Cupola Furnace &  Flow   &  0.057 & 2 \\ 
  \hline
  Pig Iron to Other &  Flow  & 0.046  &  2\\ 
  \hline
  Import to Scrap Consumption & External Input & 3.72 & 2 \\
  \hline
  Purchased Scrap to Scrap Collected & External Input & 70.98 & 2\\
  \hline
  Scrap Collected to Export &  Flow  &  21.4  &  2\\ 
  \hline
  Scrap Consumption to Blast Furnace &  Flow  & 2.62  & 2 \\ 
  \hline
  Scrap Consumption to Basic Oxygen Furnace &  Flow  & 8.35   & 2 \\
  \hline
  Scrap Consumption to Electric Arc Furnace &  Flow   & 50.9  & 2 \\
  \hline
  Scrap Consumption to Cupola Furnace &  Flow  &  1.11 & 2 \\ 
  \hline
  Scrap Consumption to Other &  Flow  &  0.167 & 2 \\ 
  \hline
  BOF\_CC to Continuous Casting &   Flow & 36.281  & 4\\
  \hline
  HSM\_Yield to Hot Rolled Sheet &  Flow  & 19.544   & 3 \\ 
  \hline
  CRM\_Yield to Cold Rolled Sheet &   Flow  & 11.079  &  3\\ 
  \hline
  Plate Mill to Plates &  Flow  &  9.12 & 3 \\ 
   \hline
  RBM\_Yield to Reinforcing Bars &  Flow  & 5.65  & 3 \\ 
  \hline
  RBM\_Yield to Bars &  Flow  &  6.7 & 3\\ 
   \hline
  RBM\_Yield to Wire and Wire Rods &  Flow  & 2.784  & 3 \\ 
  \hline
  RBM\_Yield to Light Section &  Flow  & 2.13  &  3\\ 
 \hline
  SM\_Yield to Heavy Section &  Flow  &  5.03 &  3\\ 
  \hline
  SM\_Yield to Rail and Rail Accessories &  Flow  & 1.009  & 3 \\
  \hline
  PM\_Yield to Export &  Flow  & 0.817  & 3 \\
  \hline
  Tin Mill to Tin Mill Products &  Flow  & 2.009  &  3\\
  \hline
  Galvanized Plant to Galvanized Sheet &  Flow  &  16.749 &  3\\
  \hline
  Pipe Welding Plant to Pipe and Tubing &  Flow  & 2.165  & 3\\
  \hline
  Seamless Tube Plant to Pipe and Tubing &  Flow  & 2.162 & 3 \\
  \hline
  Cold Rolled Sheet to Automotive & Ratio  & 0.25  & 5\\
  \hline
  Cold Rolled Sheet to Machinery & Ratio  & 0.079  & 5\\
  \hline
  Cold Rolled Sheet to Steel Products & Ratio & 0.313  & 5\\
  \hline
  Cold Rolled Sheet to Export & Ratio & 0.112 & 5\\
  \hline
  Galvanized Sheet to Construction & Ratio & 0.19  & 5\\
  \hline
  Galvanized Sheet to Automotive & Ratio & 0.42  & 5\\
  \hline
  Galvanized Sheet to Export  & Ratio  &  0.15 & 5\\
  \hline
  Hot Rolled Sheet to Construction & Ratio & 0.59  & 5\\
  \hline
  Hot Rolled Sheet to Automotive & Ratio & 0.133  & 5\\
  \hline
  Hot Rolled Sheet to Machinery  & Ratio &  0.108  & 5\\
  \hline
   Hot Rolled Sheet to Energy & Ratio & 0.01  & 5\\
  \hline
  Hot Rolled Sheet to Steel Products & Ratio & 0.0027  & 5\\
  \hline
  Hot Rolled Sheet to Export  &  Ratio & 0.065  & 5 \\
  \hline
  Pipe and Tubing to Construction & Ratio & 0.227  & 5\\
  \hline
  Pipe and Tubing to Automotive  & Ratio &  0.08 & 5\\
  \hline
  Pipe and Tubing to Machinery & Ratio & 0.04  & 5 \\
  \hline
  Pipe and Tubing to Energy & Ratio & 0.55  & 5\\
  \hline
  Pipe and Tubing to Export  & Ratio & 0.065  & 5\\
  \hline
  Plates to Construction & Ratio &  0.0408 & 5\\
  \hline
  Plates to Automotive  & Ratio &  0.01 & 5\\
  \hline
  Plates to Machinery & Ratio & 0.5187  & 5\\
  \hline
  Plates to Energy & Ratio & 0.067  & 5\\
  \hline
  Plates to Export  & Ratio &  0.231  & 5\\
  \hline
  Bars to Construction &  Ratio & 0.152 &  5\\
  \hline
  Bars to Automotive  & Ratio & 0.311  & 5\\
  \hline
  Bars to Machinery & Ratio &  0.238 & 5\\
  \hline
  Bars to Energy & Ratio & 0.046  & 5\\
  \hline
  Bars to Export  & Ratio & 0.131  & 5\\
  \hline
  Reinforcing Bars to Construction & Ratio & 0.925  & 5 \\
  \hline
  Reinforcing Bars to Export       & Ratio &  0.039 & 5 \\
  \hline
  Tin Mill Products to Automotive  & Ratio & 0.006  & 5 \\
  \hline
  Tin Mill Products to Steel Products & Ratio & 0.685  & 5\\
  \hline
  Tin Mill Products to Export       & Ratio &  0.067 & 5\\
  \hline
  Wire and Wire Rods to Construction & Ratio & 0.388  & 5\\
  \hline
  Wire and Wire Rods to Automotive  & Ratio & 0.285 & 5\\
  \hline
  Wire and Wire Rods to Machinery & Ratio & 0.1  &  5\\
  \hline
  Wire and Wire Rods to Energy & Ratio &  0.049 & 5\\
  \hline
  Wire and Wire Rods to Export  & Ratio & 0.094  & 5\\
  \hline
  Rail and Rail Accessories to Construction & Ratio &  0.779 & 5\\
  \hline
  Rail and Rail Accessories to Machinery & Ratio & 0.047  & 5 \\
  \hline
  Rail and Rail Accessories to Export  & Ratio & 0.141 & 5 \\
  \hline
  Light Section to Construction & Ratio &  0.86 & 5 \\
  \hline
  Light Section to Automotive & Ratio & 0.026  & 5\\
  \hline
  Light Section to Export  & Ratio & 0.057  & 5\\
 \hline
  Heavy Section to Construction & Ratio &  0.877 & 5\\
  \hline
  Heavy Section to Export & Ratio & 0.092  & 5\\
  \hline
  Steel Product Casting to Construction & Ratio & 0.259  & 5\\
  \hline
  Steel Product Casting to Automotive  & Ratio & 0.385  & 5\\
 \hline
  Steel Product Casting to Machinery & Ratio & 0.259  & 5\\
  \hline
  Steel Product Casting to Export &  Ratio & 0.111  & 5\\
  \hline
  Iron Product Casting to Construction & Ratio & 0.311  & 5\\
  \hline
  Iron Product Casting to Automotive  & Ratio & 0.552  & 5 \\
 \hline
  Iron Product Casting to Machinery & Ratio & 0.066  & 5 \\
  \hline
  Iron Product Casting to Export & Ratio & 0.07  & 5\\
  \hline
  \caption{MFA data from 2012.}
 \end{longtable}
\label{table: 2012 data}
\end{center}

\begin{table}[hb]
\begin{tabular}{cc}
 \hline
 \multicolumn{2}{l} {\textbf{Reference in Table 3}} \\
 \hline\hline
 1  & \makecell[l]{USGS. 2012. Iron Ore. Minerals Yearbook.  https://www.usgs.gov/centers/ \\ national-minerals-information-center/iron-ore-statistics-and-information}\\
 \hline
 2 & \makecell[l]{USGS. 2012. Iron and Steel Scrap. Minerals Yearbook. https://www.usgs.gov/centers/\\national-minerals-information-center/iron-and-steel-scrap-statistics-and-information} \\
  \hline
 3 & \makecell[l]{USGS. 2012. Iron and Steel. Minerals Yearbook. https://www.usgs.gov/centers/\\national-minerals-information-center/iron-and-steel-statistics-and-information} \\
  \hline
 4 & \makecell[l]{WorldSteel. 2017. Steel Statistical Yearbook 2017. https://worldsteel.org/steel-by-topic/\\statistics/steel-statistical-yearbook/} \\
  \hline
 5 & \makecell[l]{Yongxian Zhu, Kyle Syndergaard, and Daniel R. Cooper. Environmental Science \\ \& Technology  2019 53 (19) 11260-11268. DOI: 10.1021/acs.est.9b01016}\\
  \hline
 \end{tabular}
\end{table}

\bibliography{supplement}
\bibliographystyle{unsrturl}